\documentclass[12pt]{article}
\usepackage{amsfonts,amsmath}
 \usepackage{amssymb}
 \usepackage[hypertex] {hyperref}

\makeatletter \@addtoreset{equation}{section} \makeatother

\newcommand{\dr}{{\rm d}}
\newcommand{\be}{\begin{equation}}
\newcommand{\ee}{\end{equation}}
\newcommand{\bee}{\begin{eqnarray}}
\newcommand{\beee}{\begin{array}}
\newcommand{\eee}{\end{eqnarray}}
\newcommand{\eeee}{\end{array}}

\newcommand{\vac}{\pi}

\newcommand{\un}{{\underline{n}}}
\newcommand{\um}{{\underline{m}}}

\newcommand{\ga}{\alpha}

\newcommand{\gb}{\beta}
\newcommand{\gga}{\gamma}

\newcommand{\M}{{\cal M}}
\newcommand{\I}{{\cal I}}

\newcommand{\N}{{\cal N}}
\newcommand{\W}{{\cal W}}

\newcommand{\F}{{\cal F}}
\newcommand{\Ll}{{\cal L}}

\newcommand{\rhs}{{\it r.h.s.} }

\newcommand{\ie}{{\it i.e.,} }
\newcommand{\ls}{\!\!\!\!\!\!}

\newcommand{\gd}{\delta}

\newcommand{\gvep}{\varepsilon}

\newcommand{\gs}{\sigma}

\newcommand{\go}{\omega}

\newcommand{\q}{\,,\qquad}

\newcommand{\dga}{{\dot{\alpha}}}
\newcommand{\dgb}{{\dot{\beta}}}

\newcommand{\nn}{\nonumber}

\newcommand{\half}{\frac{1}{2}}

\newcommand{\p}{\partial}

\newcommand{\f}{\frac}
\newcommand{\B}{{\cal B}}
\newcommand{\C}{{\cal C}}
\newcommand{\R}{{\cal R}}

\newcommand{\U}{\Upsilon}

\begin{document}

\begin{flushright}
{\small FIAN/TD/02-18}
\end{flushright}
\vspace{1.7 cm}

\begin{center}
{\large\bf From  Coxeter Higher-Spin Theories to
\vspace{0.2cm}
 Strings and Tensor Models}

\vspace{1 cm}

{\bf  M.A.~Vasiliev}\\
\vspace{0.5 cm}
{\it
 I.E. Tamm Department of Theoretical Physics, Lebedev Physical Institute,\\
Leninsky prospect 53, 119991, Moscow, Russia}

\end{center}

\vspace {1cm}

{\it $\phantom{MMMMMMMMMMMMMMMMMMMMMMMMMMMM}$ To my father}


\vspace{0.4 cm}

\vspace{0.4 cm}

\begin{abstract}
\noindent
A new class of higher-spin gauge theories associated with various
Coxeter groups is proposed. The emphasize  is on the $B_p$--models.
The cases of $B_1$ and its infinite graded-symmetric product
$sym\,(\times B_1)^\infty$ correspond to the usual
higher-spin theory and its multi-particle extension, respectively. The multi-particle $B_2$--higher-spin theory
is conjectured to be associated with String Theory. $B_p$--higher-spin
models with $p>2$ are anticipated to be dual to the rank-$p$ boundary tensor sigma-models.
$B_p$ higher-spin models with $p\geq 2$   possess two  coupling constants
responsible
for higher-spin interactions in  $AdS$ background and stringy/tensor effects, respectively.
The brane-like idempotent extension of the Coxeter higher-spin theory is proposed allowing to unify
in the same model the fields supported by space-times of different dimensions.
Consistency of the holographic interpretation of
the boundary matrix-like model in the  $B_2$-higher-spin model is shown
to demand $N\geq 4$ SUSY, suggesting duality with the $N=4$ SYM
upon spontaneous breaking of higher-spin symmetries.
The proposed models are shown to admit unitary truncations.

\end{abstract}

\textheight 22.2 true cm

\newpage

\tableofcontents

\newpage

\textheight 21.7 true cm
\section{Introduction}
\label{intro}
Higher-spin (HS) gauge theories, that describe interactions of massless
fields of all spins, provide an interesting arena for testing
principles  of fundamental physics and, in particular, the
$AdS/CFT$ correspondence conjecture
\cite{Maldacena:1997re,Gubser:1998bc,Witten:1998qj}.
 First example of fully nonlinear HS theory
was  given for the $4d$ case in \cite{con}, while its
modern formulation was worked out  in \cite{more} (see \cite{Vasiliev:1999ba}
for a review). A specific property of HS gauge theories is that
consistent interactions of propagating massless fields demand
a curved background  providing a length scale in HS  interactions that contain
higher derivatives. $(A)dS$ is the most symmetric curved background compatible
with HS interactions \cite{Fradkin:1987ks}. The $AdS_4$ HS
model is the simplest nontrivial one in the sense that $d=4$ is the
lowest dimension where HS massless fields propagate.

There were several long-standing problems in HS theories. One was the issue of locality
raised originally in \cite{prok,Prokushkin:1999xq} where it was shown that current interactions in
HS theories can be removed by a seemingly local field redefinition containing
infinite power series in higher derivatives with the coefficients containing inverse powers
of the cosmological constant. This issue has been further analyzed in a number of papers
\cite{Vasiliev:2015wma}-\cite{Ponomarev:2017qab}. The conclusion of \cite{Vasiliev:2016xui,Vasiliev:2017cae}
was that unfolded HS equations lead to correct local interactions in the lowest order. These results
are now extended to higher orders due to the development of the appropriate homotopy technics
in \cite{GV,DGKV} where it is explained in particular how to identify the
minimally nonlocal frame of \cite{Vasiliev:2017cae} and to decrease the  level of
nonlocality in HS equations of \cite{more} in higher orders.

Once the  problem of locality in HS gauge theory has been sorted out,
the  fundamental remaining question is how to construct more general HS models that
could be related to String Theory \cite{Green:1987sp}. This is the problem addressed in this paper where we
propose a new class of  HS-like gauge theories that contain  richer spectra of fields
than the standard models of \cite{more,prok}. The proposed models are argued to be
rich enough to lead to  String Theory-like
models with massive HS fields via spontaneous breakdown of HS symmetries.
A concrete proposal for the
realization of this idea briefly discussed in the end of this paper can shed light
on the origin of  the most symmetric phase of String Theory discussed long ago in
\cite{Gross:1987ar,Gross:1988ue}. (For related discussion see also
\cite{Vasiliev:1987zv}-\cite{Sagnotti:2013bha}.)
Although this conjecture  was supported by the analysis of
high-energy limit of string amplitudes \cite{Gross:1988ue} and passed some
nontrivial tests \cite{Bianchi:2003wx}-\cite{Bianchi:2005yh},
no satisfactory understanding of this relation beyond the free field sector
of the tensionless limit of String Theory
\cite{Lindstrom:2003mg,Bonelli:2003kh,Sagnotti:2003qa} was available.
Note however that an interesting  idea of the singleton string whose spectrum is represented
by multiple tensor products of singletons was put forward  in
\cite{Engquist:2005yt,Engquist:2007pr}.

 The idea of our construction came from the
recent development of the $AdS/CFT$ holography \cite{Maldacena:1997re,Gubser:1998bc,Witten:1998qj}.
Indeed,  the fact that HS theories are most naturally formulated in the $AdS$ background
was conjectured  to play a role in the context of
$AdS/CFT$ correspondence \cite{Konstein:2000bi}-\cite{Sezgin:2002rt}.
This expectation conforms to the fundamental
result of Flato and Fronsdal \cite{FF} on the relation between
tensor products of $3d$ conformal fields (singletons) and
infinite towers of $4d$ massless fields that appear in HS theories.

In the important
work of Klebanov and Polyakov \cite{Klebanov:2002ja} it was
argued that the HS gauge theory of \cite{more} should be
dual to the $3d$ $O(N)$ sigma-model in the $N\to \infty$ limit.
The Klebanov-Polyakov conjecture was checked
by Giombi and Yin in \cite{Giombi:2009wh,Giombi:2010vg}
where it was shown in particular how the bulk computation in
HS gauge theory reproduces at least some
of conformal correlators in the free $3d$ theory. (See also important papers
\cite{Maldacena:2011jn,Maldacena:2012sf}.)

Original Klebanov-Polyakov conjecture \cite{Klebanov:2002ja} related large-$N$ limit of the $3d$
boundary sigma-model with the action
\be
S = \half\int d^3x \big( \p_\nu \varphi_i \p^\nu \varphi^i + \f{\lambda}{N}(\varphi_i  \varphi^i)^2\big )
\ee
for scalar fields $\varphi^i$ in the vector representation of $O(N)$
($i=1,2,\ldots, N$) with the particular  HS gauge theories of \cite{more},
so-called $A$ and $B$-models,  in the bulk.
In \cite{Aharony:2011jz} Klebanov-Polyakov conjecture was
extended to the boundary vector models with Chern-Simons interaction where
the Chern-Simons level was related to the free phase parameter in the models
of \cite{more}. Fermionic analogues of these conjectures were put forward in
\cite{Leigh:2003gk,Sezgin:2003pt,Giombi:2011kc} and were partially
checked in \cite{Giombi:2012ms} (for more recent results see
\cite{Sezgin:2017jgm,Didenko:2017lsn}), suggesting a  nontrivial duality  between bosonic and fermionic boundary
models, so-called bosonization, \cite{Jain:2013gza,Gur-Ari:2015pca,Seiberg:2016gmd}.

Some difficulties of  the original analysis of  HS holography originated from non-locality
of the  naive perturbative approach in HS theory leading to
 divergencies in the holographic tests \cite{Giombi:2009wh}.
 In \cite{Sezgin:2017jgm,Didenko:2017lsn,Misuna:2017bjb} it
has been checked that unfolded HS equations in the local frame of
\cite{Vasiliev:2016xui,Gelfond:2017wrh,Vasiliev:2017cae}
properly reproduce the anticipated
holographic results including the  general case of Chern-Simons boundary
theory, thus opening a way towards further analysis of HS holography in
the framework of the nonlinear HS system of \cite{more}.

$AdS_4/CFT_3$ HS duality was extended to
$AdS_3/CFT_2$ holography by Gaberdiel and Gopakumar \cite{Gaberdiel:2010pz}
(see \cite{Gaberdiel:2012uj} for a review) that has been extensively
studied in the literature (see e.g. \cite{Henneaux:2010xg}-\cite{Beccaria:2013wqa}).
Elaborating on $AdS_3/CFT_2$ duality,  Gaberdiel and Gopakumar arrived at
the interesting conjecture on the relation between certain compactifications
of String Theory and $3d$ HS theories \cite{Gaberdiel:2014cha}
as well as on the structure of stringy HS symmetry
\cite{Gaberdiel:2015mra}, further elaborated in
\cite{Gaberdiel:2015uca}-\cite{Gaberdiel:2018rqv} (see also
\cite{Giribet:2018ada}).
Remarkably, symmetries of the string-like HS theories proposed in this paper exhibit
interesting similarities with the Gaberdiel-Gopakumar construction.

The vector boundary models were recently  extended
to so-called tensor
models \cite{Klebanov:2016xxf}
with the scalar and spinor  fields $\varphi_{i_1 i_2 \ldots i_p}$
carrying several color indices (symmetrized
and/or traceless or not).   The Lagrangian for the so-called $O(N)^3$ boundary model is
\be
\label{tact}
S = \half\int d^dx \big( \p_\un \varphi_{ijk} \p^\un \varphi^{ijk} + g \varphi_{ijk}
\varphi^i{}_{nm}  \varphi^{jn}{}_l\varphi^{kml}\big )\,.
\ee
For $O(N)^p$ tensor models with any $p>1$ the interaction vertex is of order
$p+1$. One can also consider bosonic and fermionic tensor models of $U(N)^p$ and
$USP(2N)^p$ types. Recently, these models have been extensively studied in the
literature \cite{Beccaria:2017aqc}-\cite{Bulycheva:2017ilt}
in particular due to their relation \cite{Witten:2016iux} to the SYK model
\cite{Sachdev:1992fk,K},
 providing a new class of models possessing an interesting large $N$ regime
 \cite{Maldacena:2016hyu}-\cite{Gross:2017aos}.
This suggests in turn that they should admit some holographically dual description.
The case of $p=2$ corresponds to models with matrix-valued fields like in $N=4$
SYM theory. In this case, the holographically dual theory is String Theory.

To the best of our knowledge no models holographically dual to the tensor
sigma-models with $p>2$ beyond the original one-dimensional SYK model
\cite{Sachdev:1992fk}-\cite{Maldacena:2016hyu}
 are available in the literature though  ideas that there
should be some HS holographic description analogous to that for the $O(N)$
model were expressed e.g. in \cite{Gross:2017aos,deMelloKoch:2017bvv}. It is immediately clear that the HS
model of \cite{more} as well as its generalizations to three \cite{prok} and any \cite{Vasiliev:2003ev}
 dimensions are not appropriate duals since the spectrum of $O(N)$
singlet operators  of rank-$p$ tensor models increases with $p$  due to the
presence of new multi-particle states \cite{Beccaria:2017aqc,Bulycheva:2017ilt}. The same time, the sigma-model form of the boundary theory
suggests that a bulk dual model has to be of a HS type.

An obvious class of conformal
boundary models holographically dual to tensorial HS theories in the bulk
first discussed by Becaria and Tseytlin in \cite{Beccaria:2017aqc} is provided by
the free boundary conformal
fields $\varphi_{i_1,i_2,\ldots i_n}$  with the tensorial currents of the type of the vertex in
(\ref{tact}), \ie
$J =\varphi_{ijk} \varphi^i{}_{nm}  \varphi^{jn}{}_l\varphi^{kml}$, or similar.
Analogously to the Klebanov-Polyakov conjecture \cite{Klebanov:2002ja} for vector models, we conjecture that
the proposed Coxeter HS theories are holographically dual to such higher currents in the free boundary models.
Moreover, the $AdS_4$ Coxeter HS models admit an extension to complex coupling constants that should be
holographically dual to the $3d$ boundary conformal models with Chern-Simons interactions analogous to the models
considered
in \cite{Aharony:2011jz} for the  vectorial HS holography.
A subtlety of the holographic interpretation of the boundary tensor sigma-models
is that their conformal structure analogous to that of critical sigma-model
in the Klebanov-Polyakov conjecture \cite{Klebanov:2002ja} is not yet obvious. Note however
 that, as argued in \cite{Vasiliev:2012vf}, it is not {\it a priory}
guaranteed that the boundary duals
of the class of HS theories proposed in this paper are represented by local
boundary sigma-models free from additional interactions via boundary HS conformal
fields  which may effectively induce nonlocal interactions
on the boundary  (see also \cite{Giombi:2013yva}).

An important problem in the context of HS duality consists of the construction of so-called multi-particle HS theories
containing multi-particle composite operators of the original HS theory as fundamental operators.
In \cite{Vasiliev:2012tv} a multi-particle extension of the usual HS symmetry was proposed, conjectured
to underly an infinite extension of the conventional HS theory.
 The aim of this paper is  to propose
a class of generalized nonlinear HS systems  to be associated with
the tensor boundary sigma-models and their multi-particle extensions including the vector
and tensor ones.

The proposed models are characterized by an integer
$p$ analogous to the rank of the tensor  $\varphi_{i_1 i_2 \ldots i_p}$ and have
a much richer spectrum than usual HS theory of \cite{more,prok,Vasiliev:2003ev}. They
are based on  the deformed oscillator algebras found in
\cite{Polychronakos:1992zk,Brink:1992xr,Brink:1993sz} in
the context of $p$-particle Calogero models,  known as
Cherednik algebras \cite{Cherednik}, which in turn are associated with
various Coxeter groups. HS-like models of this class
could  be  formulated long ago and potential relevance of
the Cherednik algebra to HS theory was mentioned already in
\cite{Brink:1993sz}. However, naive extension of this class
 was not formulated explicitly so far because the resulting
 spectra of states left no room for a massless spin-two state, \ie graviton,
 not allowing the resulting models to fit  into the standard paradigm of HS gravity.
The same time, the naive extension to HS theories of this kind exhibits  interesting features
indicating their potential relevance as bulk duals of the tensor boundary sigma-models.

In this paper we extend the construction of Cherednik algebras to {\it framed
Cherednik algebras} in such a way that the related HS theories turn out to be free
of the aforementioned difficulties, containing massless HS fields in the spectrum.
This generalization was inspired by the construction of multi-particle algebras
of \cite{Vasiliev:2012tv} and, other way around, allows us to formulate nonlinear
equations for the multi-particle HS theory.
It is based on the $q^{th}$ direct product degree of a chosen rank-$p$ Coxeter
group $\C$, thus containing two positive integers as free parameters:
 $p$ and  $q$ are associated, respectively, with the tensor degree of the boundary model and
the degree of multi-particle states. For instance
 $\C=A_1\sim B_1$ and $q=1$ or $q=\infty$ correspond to the conventional HS theory or
 its multi-particle extension, respectively. On the other hand, higher
ranks $p$ of the Coxeter group $B_p$ are conjectured to be associated with the rank-$p$
tensor boundary models. Generalization to other Coxeter groups and root systems
(for more  detail on Coxeter groups see  for instance \cite{burb}) is also possible
and worth to be investigated. The case of $p=2, q=\infty$ is particularly appealing in connection
with the HS interpretation of String Theory. The $B_2$-model possesses two  coupling
constants responsible for HS interactions in  $AdS$ background and stringy
effects, respectively. It is tempting to speculate that this model can provide
a realization of the most symmetric phase of String Theory.
A  possible HS symmetry breaking mechanism is also discussed
in the concluding section of this paper.

 The main idea of this work was to look for  HS bulk models providing generalizations
of the usual HS theory to models with richer spectrum that would qualitatively
 match the pattern of operators of the tensor boundary models and their multi-particle
 extensions.
 It should be stressed that the formal consistency  and gauge symmetries
 along with simple physical conditions like, {\it e.g.}, Lorentz covariance severely
 restrict possible higher-rank extensions of HS theories. The  HS-like
 models based on framed Cherednik algebras provide a  distinguished and, may be,
 exhaustive possibility for the
 construction of models fulfilling these properties.

The paper is organized as follows. We start with recalling
the form of standard nonlinear HS equations  in Section \ref{Standard formulation}.
In Section \ref{Deformed oscillator algebras} we recall the construction of deformed
oscillators that underlies the standard formulation of HS equations as well as its
Cherednik extension associated with an arbitrary Coxeter root system. In Section
\ref{hsmod} we explain the difficulty of the naive extension of HS algebras via
enlargement of the number of species of oscillators and introduce the notion of the
{\it framed } oscillator algebra free of this difficulty. Then in Section \ref{FCA}
we apply this idea to the construction of framed Cherednik algebras underlying the
proper extension of HS equations to any Coxeter system presented in Section \ref{CHS}.
Further extensions of the proposed systems to higher differential forms generating
invariant functionals, color systems with Chan-Paton factors, and idempotents associated
with brane-like dynamical systems in different dimensions   are presented in Sections
\ref{further}, \ref{Matrix} and \ref{Proj}, respectively.
 Unitarity of the proposed models
is discussed in Section \ref{unitarity}. Graded extension of
the construction of multi-particle algebras of  \cite{Vasiliev:2012tv} along with the discussion
of its different frames and new idempotent construction are presented in Section \ref{frames}.
General properties
of the proposed systems, their interpretation in the context of string-like and
tensor-like HS systems and a  possible HS symmetry breaking mechanism are discussed
in Section \ref{interpretation} where it is also argued that
 the proposed scheme allows one to interpret composite fields in the original
space-time as elementary fields in  higher dimensions. Section \ref{concl} contains brief conclusions.

\section{Standard higher-spin equations}
\label{Standard formulation}

Nonlinear HS equations of \cite{more,prok,Vasiliev:2003ev} are formulated in terms
of three types of fields
\be
\label{fields}
 W(Z;Y;k|x)=dx^\un W_\un (Z;Y;k|x)\q B(Z;Y;k|x)\,,
 \qquad S=dZ^A S_A (Z;Y;k|x)\,.
\ee
Here $x^\un$ are space-time coordinates while the variables $Z^A$ and
$Y^A$ ($A,B,\ldots = 1,\ldots M$) are
auxiliary coordinates acquiring different interpretation in the spinorial
HS models of \cite{more,prok} and vectorial HS model of \cite{Vasiliev:2003ev}.
$dZ^A$ are anticommuting  $Z-$differentials
\be
 dZ^A dZ^B=-dZ^B
dZ^A\q dZ^A dx^\un = -dx^\un
dZ^A\q dx^\um dx^\un = -dx^\un
dx^\um\,.
\ee

 HS equations of \cite{more,prok,Vasiliev:2003ev} have the form
\be
\label{dW}
\dr W+W*W=0\,,\qquad
\ee
\be
\label{dB}
\dr B+W*B-B*W=0\,,\qquad
\ee
\be
\label{dS}
\dr S+W*S+S*W=0\,,
\ee
\be
\label{SB}
S*B=B*S\,,
\ee
\be
\label{SS}
S*S= -i (dZ^A dZ_A + (dz^\ga dz_\ga  F_*(B) *\kappa + c.c.) )
\,.
\ee
Here $\dr=dx^\un \f{\p}{\p dx^\un}$ is the space-time de Rham differential and $*$ denotes a model-dependent star product.
$F_*(B) $ is some star-product function of the field $B$.
Index $\ga$ takes two values enumerating basis elements of some subspace of the space
labelled by $A$. The complex conjugated term $c.c.$ does not contribute
in the real models of \cite{prok,Vasiliev:2003ev} but has to be added
in the $4d$ model of \cite{more}. $\kappa$ is an element of the algebra
defined to commute with everything except for $dz^\ga$ with which it anticommutes
\be
\label{kdz}
\kappa * dz^\ga = - dz^\ga *\kappa\,.
\ee

The associative star product $*$ acts on functions of two
spinor variables
\be
\label{star2}
(f*g)(Z;Y)=\frac{1}{(2\pi)^{4}}
\int d^{4} U\,d^{4} V \exp{[iU^A V^B C_{AB}]}\, f(Z+U;Y+U)
g(Z-V;Y+V) \,,
\ee
where
$C_{AB}$ is a symplectic form and
$ U^A $, $ V^B $ are real integration variables. It is
normalized in such a way that 1 is a unit element of the star-product
algebra, \ie $f*1 = 1*f =f\,.$ Star product
(\ref{star2}) provides a particular
realization of the Weyl algebra
\be
\label{YZ}
[Y_A,Y_B]_*=-[Z_A,Z_B ]_*=2iC_{AB}\q
[Y_A,Z_B]_*=0
\ee
($[a,b]_*=a*b-b*a$) resulting from the normal ordering
 with respect to the elements
\be
\label{oscrel}
   b_A = \frac{1}{2i} (Y_A - Z_A )\,,\qquad
   a_A = \frac{1}{2} (Y_A + Z_A )\,,\qquad
\ee
which satisfy
\be
\label{com a}
   [a_A, a_B]_*=[b_A, b_B]_* =0 \,,\quad
   [a_A, b_B]_* =C_{AB} \,.
\ee
The choice of the star-product realization (\ref{star2}) of the Weyl algebra is significant
from many perspectives and, in particular, for the analysis of locality in \cite{GV,DGKV}.
An important property of the star product (\ref{star2}) is that it
admits the inner Klein operator
\be
\label{vark}
\varkappa = \exp i Z_\ga Y^\ga \,,
\ee
which behaves as $(-1)^{N} ,$ where $N$ is the oscillator number
operator, \ie
\renewcommand{\U}{\varkappa}
\be
\U *\U =1,
\ee
\be
\label{[UF]}
\U *f(Z;Y)=f(\tilde Z;\tilde Y)*\U\,,\quad
\ee
where $\tilde Z^A = (-Z^\ga, Z^a)$  for any $Z^A=(Z^\ga, Z^a)$ with $Z^a$ being
the rest components of $Z^A$. The Klein operator
\be
k= \kappa \U\,,
\ee
 on which all fields (\ref{fields}) depend, anticommutes with $Y^\ga$, $Z^\ga$ and $dZ^\ga$.
 Hence,  $\kappa$ in Eq.~(\ref{SS}) is represented as
\be
 \kappa=k \U\,.
\ee
(Note that the fields of the $4d$ model of \cite{more} also depend on the conjugated Klein
operator $\bar k$.)

\section{Deformed oscillator algebras}
\label{Deformed oscillator algebras}

\subsection{Single oscillator}
\label{single}

The nontrivial part of the HS equations is represented by Eqs.~(\ref{SB}), (\ref{SS}).
Since $B$ is covariantly constant by (\ref{dB}), and commutes with $S$ by (\ref{SB})
and   $\kappa$ since it is a $dz$-independent zero-form,
it behaves as a central element. Replacing $F_*(B)$
by a central element $\nu$ and discarding for a moment the complex conjugated term
we observe that the nontrivial part of the HS equations has the form
\be
\label{SSnu}
S*S= -i (dZ^A dZ_A + \nu * dz^\ga dz_\ga  *\kappa  )\q \ga=1,2\,.
\ee

This equation has many important interpretations for particular realizations
of $S$, $*$ and $\kappa$ (see e.g.
\cite{Vasiliev:1999ba,Vasiliev:2015wma}).
The most important  property in the context of this paper is that the \rhs
of (\ref{SSnu}) commutes with $S$ despite of (\ref{kdz}) since any trilinear combination of
$dz^\ga$  is zero because $\ga=1,2$ and $dz^\ga$  anticommute.
It is this property that guarantees formal consistency of HS equations (\ref{dW})-(\ref{SS}).

 Eq.~(\ref{SSnu}) induces commutation relations of
the deformed oscillator algebra. Indeed, let  $S=dz^\ga q_\ga +\ldots$,
where ellipses denotes terms proportional to components of $dZ^A$ different from
$dz^\ga$, and
\be
\tilde \nu = \nu k\,,
\ee
where
\be
k dz^\ga = - dz^\ga k\q k q_\ga = - q_\ga k\q k^2 =1\,.
\ee
So defined $k$ and $\tilde \nu$ still commute with $S$ while, setting
\be
\varkappa := \kappa k\,,
\ee
(\ref{SSnu}) implies deformed oscillator commutation relations in the form of \cite{Aq}
\be
\label{defo}
[q_\ga\,,q_\gb] = -2i\epsilon_{\ga\gb} (1 +\tilde \nu \varkappa)\q
\varkappa q_\ga = - q_\ga\varkappa\q \varkappa^2=1\,,
\ee
where $\tilde \nu$ is central. The enveloping algebra $Aq(\tilde \nu)$
of these relations,
which is the algebra of polynomials $f(q,\varkappa)$ with $\tilde \nu$ being a
parameter, has the fundamental property \cite{Aq} that
\be
t_{\ga\gb}:= \f{i}{4}\{q_\ga\,,q_\gb\}
\ee
generate $sp(2)$
\be
\label{tt}
[t_{\ga\gb}\,, t_{\gga\gd}] = \epsilon_{\gb\gga}t_{\ga\gd}  + \epsilon_{\gb\gd}t_{\ga\gga} + \epsilon_{\ga\gga}t_{\gb\gd} + \epsilon_{\ga\gd}t_{\gb\gga}
\ee
rotating $q_\ga$ as a $sp(2)$ vector
\be
[t_{\ga\gb}\,, q_{\gga}] = \epsilon_{\gb\gga}q_{\ga} +\epsilon_{\ga\gga}q_{\gb}\,.
\ee
By these properties, the deformed oscillators (\ref{defo}) provide
a representation-free realization of the Wigner oscillator \cite{wig}
 considered by many authors (see, {\it e.g.}, \cite{yang,deser,mukunda}).

That the deformed oscillator algebra possesses the $sp(2)$ automorphism
algebra implies that  nonlinear HS equations (\ref{dW})-(\ref{SS})
are also  $sp(2)$ covariant. In
$4d$ HS theory of \cite{more} and $3d$ HS theory of
\cite{prok} the respective $sp(2|\mathbb{C})$ and   $sp(2|\mathbb{R})$
imply usual Lorentz covariance \cite{Vasiliev:1999ba}.
 In vectorial models in any dimension of \cite{Vasiliev:2003ev},
$sp(2|\mathbb{R})$ is a Howe dual algebra allowing to truncate away extra states
 from the spectrum.

\subsection{Many oscillators}

\label{manya}

The deformed oscillator of Section (\ref{single}) can be interpreted as describing the
relative motion  in the two-body Calogero model. Extension
 to the $\N$-body Calogero model  proposed in
\cite{Polychronakos:1992zk,Brink:1992xr,Brink:1993sz}
consist of $\N$ copies of oscillators
$q_{\ga\,i} $ with $\ga=1,2$, $i=1,\ldots \N$ and the generators $K_{ij}$ of
the symmetric group $S_\N$ that obey
\be
\label{Kq}
K_{ij}q_{\ga\, i } = q_{\ga j} K_{ij}\,,
\ee
\be
\label{sym}
K_{ij}= K_{ji}\q K_{ij} K_{ij} = I\q K_{ij} K_{jl} = K_{il}K_{ij}=K_{jl} K_{il}\,
\ee
(no summation over repeated indices),
and $K_{ij}$ commutes with $q_{\ga\, l}$ with $i\neq l$, $j\neq l$. Commutation
relations of the oscillators $q_{\ga\, l}$ have the form
\cite{Polychronakos:1992zk,Brink:1992xr,Brink:1993sz}
\be
\label{qq}
[q_{\ga\, n}\,, q_{\gb\, m}] = -i\epsilon_{\ga\gb} \big (\delta_{nm} \Big (2I +\nu \sum_{l=1}^\N K_{ln}\Big ) -\nu K_{nm} \big )\,.
\ee
Remarkably, this deformation still respects Jacobi identities since
antisymmetrization over any three two-component indices gives zero as is
most obvious from the Coxeter realization  presented
in  Section \ref{cher}.

From (\ref{qq}) it follows that the {\it center of mass coordinates}
\be
Q_\ga:= \N^{-\half}\sum_{n=1}^\N q_{\ga\,n}
\ee
have undeformed commutation relations  with themselves and  {\it relative coordinates}
\be
[Q_\ga\,, Q_\gb] = -2i  \epsilon_{\ga\gb}\q
[Q_\ga\,, q_{\gb\,n} - q_{\gb\,m}] = 0.
\ee

The fundamental property of relations (\ref{qq}) is that  the
operators
\be
t_{\ga\gb}:= \f{i}{4}\sum_{n=1}^\N\{q_{\ga\,n}\,,q_{\gb\,n}\}
\ee
 obey the $sp(2)$  commutation relations (\ref{tt}), properly
rotating all indices $\ga$ by virtue of
\be
[t_{\ga\gb}\,, q_{\gga\,n}] = \epsilon_{\gb\gga}q_{\ga\,n} +\epsilon_{\ga\gga}q_{\gb\,n}\,.
\ee
Clearly, $t_{\ga\gb}$ can be represented as a sum of $sp(2)$ generators
acting separately on the center of mass  and relative coordinates
\be
t_{\ga\gb}= t^{cm}_{\ga\gb}+t^{rel}_{\ga\gb}\q
t^{cm}_{\ga\gb}:= \f{i}{4}\{Q_{\ga}\,,Q_{\gb}\}\,.
\ee

At $\N=2$, there is only one relative coordinate
\be
q_\ga = \f{1}{\sqrt 2} (
q_{\ga 1} - q_{\gb 2})\,.
\ee
With $K_{12}=\varkappa$ (\ref{qq}) amounts to (\ref{defo}) which therefore describes
the relative coordinate sector of the two-body Calogero model.

\subsection{Coxeter groups and Cherednik algebras}
\label{cher}

A rank-$p$ Coxeter group $\C$ is  generated by reflections
with respect to a system of root vectors $\{v_a\}$ in a $p$-dimensional
Euclidean vector space $V$ with the scalar product  $(x,y)\in \mathbb{R} $,
$x, y \in V$ (for more detail see \cite{burb}).
 An elementary reflection associated with the root vector $v_a$ acts
on $x\in V$ as follows
\be
\R_{v_a}  x^i = x^i-2v^i_a \frac{(v_a,x)}{(v_a,v_a)}\q \R_{v_a}^2=I
\ee
(no summation over $a$). $p$ is the rank of the Coxeter root system $\{v_a\}$.
 Note that $\R_{v_a}$ changes a sign of $v_a$
\be
\label{rvv}
\R_{v_a} v_a = - v_a\,.
\ee

Let $q^n_{\ga}$ ($\ga=1,2$, $n=1;\ldots,p$)  obey Heisenberg commutation relations
\be
[q^n_{\ga}\,,q^m_\gb] = -2i\epsilon_{\ga\gb} \delta^{nm}\q q^n_\ga = \delta^{nm}q_{m\,\ga} \,.
\ee
Cherednik deformation of the semidirect product of the
Heisenberg algebra with the group algebra of $\C$ is
\be
\label{qqc}
[q^n_{\ga}\,,q^m_\gb] = -i\epsilon_{\ga\gb}
\left( 2\delta^{nm}+\sum_{v\in \mathcal{R}}\nu(v)\frac{v^{n}v^{m}}{(v,v)}%
K_{v}\right)\,,
\ee
\be
\label{kq}
K_{v} q^n_\ga = \R_{v}{}^n{}_m q^m_\ga K_{v}\,,
\ee
where $K_v$ generate  the Coxeter group, and the coupling constants
$\nu(v)$ are invariants of $\C$  being constant on the conjugacy classes of root
vectors under the action of $\C$. These relations generalize (\ref{qq}) from
$A_{p}=S_{p+1}$ to any Coxeter group. By definition, the operators $K_v$
obey defining relation of the Coxeter group $\C$. In particular, each $K_v$ is
involutive
\be
K_v K_v = I\,
\ee
and since the reflections generated by $v$ and $-v$ coincide, the respective reflection
generators are also identified
\be
K_{-v} = K_v\,.
\ee

 It is not difficult to check that
the double commutator of $q^n_\ga$ in (\ref{qqc}) respects the Jacobi identities
for any (not necessarily $\C$-invariant) $\nu(v)$. Indeed, the non-zero part of the triple commutator
of $q^n_\ga$, $q^m_\gb$, $q^k_\gga$ resulting from (\ref{qqc}) is proportional
$v^n v^m v^k $ and hence contains the total antisymmetrization over three two-component
indices $\ga, \gb , \gga$ giving zero. Covariance of  relations (\ref{qqc}) under
the action of the Coxeter group itself demands $\nu(v)$ be  $\C$-invariant.

In the sector of relative coordinates, the commutation relations (\ref{qq})
is a particular case of (\ref{qqc}) for the $A_{p}$ root system
\be
v^{nm} = e^n -e^m\,,
\ee
where $e^n$ form an orthonormal frame in $\mathbb{R}^{p+1}$.
$V$ is the $p$-dimensional subspace of relative coordinates in $\mathbb{R}^{p+1}$
spanned by $v^{nm}$.

Another important case of the Coxeter root system
is $B_p$ with the roots
\be
\label{con2}
R_1=\{\pm e^n \qquad 1\leq n\leq p\}\q R_2=\{ \pm e^n \pm e^m\quad 1\leq n< m\leq p\}\,.
\ee
In addition to permutations, $B_p$ contains reflections of any basis axis in
$V=R^p$ generated by $v_\pm^{n}=\pm e^n$. In this case $R_1$ and $R_2$ form two
conjugacy classes of $B_p$.
Let $K_n$, $K_{nm}$ and $K^+_{nm}$  be associated with $e^{n}$,  $  e^n -e^m$
and $  e^n +e^m$, respectively.

The defining relations of $B_p$ are represented by  relations (\ref{sym})
for the symmetric group generators $K_{nm}$ along with
\be
K_n K_{nm} = K_{nm} K_m\q K_n K_n = I\q K^+_{nm} =K_n K_{nm}K_n
\ee
(generators $K_n$, $K_{nm}$ and $K^+_{nm}$  with pairwise different indices
commute).

The Coxeter group underlying  $3d$ HS theory of \cite{prok} is
$A_1\sim B_1$.\footnote{More precisely the Coxeter root system $A_1\sim C_1$ while the
root systems $B_p$ and $C_p$ differ by the normalization of root vectors \cite{burb}
having the same Coxeter group.}
The case of $B_2$ is also of great importance since
 it is argued in Section \ref{interpretation} to be related to the string-like models.

For any Coxeter root system the generators
\be
\label{tab}
t_{\ga\gb}:= \f{i}{4}\sum_{n=1}^p\{q_{\ga}^n\,,q_{\gb}^n\}
\ee
 obey the $sp(2)$  commutation relations (\ref{tt}), properly
rotating all indices $\ga$ by virtue of
\be
\label{tqq}
[t_{\ga\gb}\,, q_{\gga}^n] = \epsilon_{\gb\gga}q_{\ga}^n +\epsilon_{\ga\gga}q_{\gb}^n\,
\ee
as is easy to see  using (\ref{kq}) and (\ref{rvv}). This fact is of fundamental
importance for HS theories.

\section{Higher-spin  modules and framed oscillator algebras}
\label{hsmod}

A natural way to allow a richer spectrum in the model including, in particular, mixed-symmetry
massless and massive
states (the latter are known to be present in String Theory \cite{Green:1987sp}) known to allow
consistent interactions at least in the lowest orders
\cite{Metsaev:1993mj}-\cite{Metsaev:2012uy},
is to let more species of oscillators in the model: $y_\ga\to y^n_\ga$. However,
the construction of oscillator algebras with multiple species of oscillators
as well as their Cherednik deformation recollected in Section \ref{cher} cannot be directly
applied to the construction of physically acceptable HS theories for the following
reason.

Consider a HS algebra called for brevity $hs_1$ with the single set of oscillators
(\ie $p=1$). It is well known \cite{Konshtein:1988yg,Gun0} that the Fock $hs_1$-module
$F_1$ describes free boundary conformal fields  \cite{Shaynkman:2001ip,Vasiliev:2001zy}, \ie
Dirac singletons \cite{Dirac:1963ta,Fronsdal:1978vb,FF,Gunaydin:1985tc,Bergshoeff:1988jm,FerFr}.
 Their conformal
dimension is identified with the weight $h_1$ of the vacuum state in $F_1$ with respect to
the dilatation operator $D$ represented by a bilinear of oscillators in $hs_1$
\be
D |0\rangle = h_1 |0\rangle\,.
\ee
(In the $AdS$ unitary frame $D$ is replaced by the energy operator $E$. For
details on the oscillator realization of
semisimple Lie algebras and  their modules see, {\it e.g.},
\cite{Bars:1982ep}.)
Correspondingly,  lowest weight representations of the naively extended
algebras $hs_p$ built from $p$ copies of oscillators will have multiple
weights in the respective Fock vacuum $F_p$,
\be
h_p = p h_1\,,
\ee
hence  carrying
 higher lowest weights.

For $p=1$, elements of the tensor product $F_1\otimes F_1$ are interpreted
either as conserved currents on the boundary or as massless fields in the bulk as was
originally shown by Flato and Fronsdal \cite{FF}.
However for $p>1$ the lowest weights of the submodules in  $F_p\otimes F_p$
turn out to be too high to describe both conserved boundary currents and bulk massless fields.
In particular, the $p>1$ modules have no room for a massless spin-two field, \ie
graviton. That the respective theory cannot contain gravity makes it impossible
 the construction of consistent HS models
via allowing more oscillators, \ie more generating elements of the star-product
algebra. (Note that this argument does not apply to $3d$ HS theories since
in this case the graviton is topological \cite{Achucarro:1987vz,Witten:1988hc},
carrying no local degrees of freedom,  and, hence, having no nontrivial associated module
of the space-time symmetry group. As a result, the construction of $3d$ HS theories turns
out to be far less restrictive than  in $d>3$.)

Recently, this difficulty has been resolved in the construction of
multi-particle algebra  \cite{Vasiliev:2015wma} defined as the universal enveloping algebra
of the usual HS algebra, \ie $U(hs_1)$. From this definition it is clear that
HS algebra-modules  form modules over the multi-particle HS algebra as well. As observed in
\cite{Vasiliev:2015wma}, an important  difference between the construction of
multi-particle algebra and  oscillator algebras with many oscillators
is due to appearance of independent unit  elements associated with each copy of oscillators.

Namely, in the usual oscillator algebras the fundamental commutation relations are
\be
\label{osci}
[q^n_\ga\,,q^m_\gb ] = 2i \delta^{nm} \epsilon_{\ga\gb} I\,,
\ee
where $I$ is the unit element of the algebra. In the multi-particle HS algebra
these relations are replaced by
\be
[q^n_\ga\,,q^m_\gb ] = 2i \delta^{nm} \epsilon_{\ga\gb} I_n\,,
\ee
 where ``units" $I_n$  assigned to the respective species of the oscillators
obey
\be
\label{Ii}
I_n I_n = I_n\q I_n I_m = I_m I_n \,.
\ee
Such algebras will be called {\it framed oscillator algebras}. Note that the framed
oscillator algebras are nothing else as multiple tensor products of the original
associative oscillator algebra with no color indices $n$ and $m$, \ie of a single
pair of oscillators while usual oscillator algebras (\ref{osci}) with the single unit
 element are their quotients.

The same time  multi-particle HS algebra is unital being endowed with the
unit element $I$ unrelated to $I_n$. Roughly speaking, $I$ is associated with the
physical vacuum with no excited states while $I_n$ is associated with the minimal
energy state in the sector of $n$-th particles.

An important feature of the  multi-particle algebras is that their elements are
(graded)symmetrized over elementary species. In that case, the only single-particle
element built from  $I_n$ is
\be
e=\sum_{i=1}^p I_i\,.
\ee
As such $e$ becomes algebraically independent of individual $I_n$ with the
consequence that its eigenvalues may be different from multiples of those for $I_n$
which may have no sense at all.

Technically, in the framed oscillator algebras, the problem with vacuum eigenvalues
is avoided as follows. The ``units" $I_n$ form a set of idempotents allowing to
consider Fock-type modules $F_p^i$ generated from vacua $|0^i\rangle$ obeying
\be
I_j |0^i\rangle = \delta_i^j |0^i\rangle\q a_-^j|0^i\rangle=0\q
\ee
\be
a_+^j|0^i\rangle=0 \quad i\neq j\,,
\ee
where $a_\pm^i$ are the creation and annihilation operators built from $q_\ga^i$.
Clearly, the modules of the generators $t_{\ga\gb}$ formed by $F_p^i$ are equivalent
to the $p=1$ Fock module $F_1$. Note that for usual oscillator algebras this
construction is applicable to the oscillators with indices $\ga,\gb$
taking any even number of values $2M$ in which case  $t_{\ga\gb}$
(\ref{tab})  generate $sp(2M)$.

Since higher-rank HS theories are anticipated to be associated with the Cherednik-like
deformation of the oscillator algebras, to let them contain gravity it is
necessary to find a framed  extension of the Cherednik algebra having a room for
 units $I_n$.

\section{Framed Cherednik algebras}
\label{FCA}

\subsection{$A_N$}
\label{AN}
To illustrate the idea let us first consider the $A_{p-1}$ system of
Section \ref{manya}. In addition to  $q_{\ga n}$ and $K_{nm}$
with $n,m=1,\ldots p$, we
introduce elements $I_n$ that obey
\be
\label{II}
I_n I_m = I_m I_n\q I_n I_n = I_n\,,
\ee
\be
\label{Iq}
I_n q_{\ga n} =  q_{\ga n} I_n =  q_{\ga n} \q
I_n q_{\ga m} = q_{\ga m} I_n\,.
\ee

In presence of idempotents $I_n$, the proper modification of the
deformed oscillator relations consistent with (\ref{Iq}) is
\be
\label{qqpi}
[q_{\ga\, n}\,, q_{\gb\, m}] = -i\epsilon_{\ga\gb} \big (\delta_{nm}
\Big (2I_n +\nu \sum_{l=1}^p \hat K_{ln}\Big ) -\nu \hat K_{nm} \big )\,,
\ee
where
\be
\hat K_{nm} = I_n I_m K_{nm}\,.
\ee
Note that operators $\hat K_{nm}$ obey all relations of the symmetric group $S_p$
except for the second relation in (\ref{sym})  replaced by
\be
\label{hKK}
\hat K_{nm}\hat K_{nm} = I_n I_m\,.
\ee
$\hat K_{nm}$ are demanded to obey the analogs of (\ref{Kq})
\be
\label{hKq}
\hat K_{ij}q_{\ga\, i } = q_{\ga j} \hat K_{ij}\,
\ee
and to commute with all
$I_l$
\be
\label{HIK}
I_l \hat K_{nm}= \hat K_{nm}I_l \quad \forall \,\,l,n,m\,.
\ee
(Note that by (\ref{II}) this is consistent with the naive
expectation that $K_{nm} I_m = I_n K_{nm}$.) Also, the following relations held
true
\be
I_n \hat K_{nm}=I_m \hat K_{nm}= \hat K_{nm}\,.
\ee

It should be stressed that the unhatted generators of the symmetric group $K_{nm}$ do not appear
in the construction of the framed Cherednik algebra. Relation (\ref{hKK})
is not of the braid group since $I_n I_m$ is not invertible. Hence, $\hat K_{nm}$
do not generate a group algebra.

This modification guarantees consistency  of  defining relations (\ref{qqpi}).
Namely,   (\ref{HIK}) implies consistency of (\ref{qqpi}) and (\ref{Iq}) while
the modification  (\ref{hKK}) of the symmetric group relations plays no role in the
Jacobi identity check
\be
[q_{n\ga}\,,[q_{m\gb}\,,q_{k\gga}]] + cycle =0\,.
\ee

\subsection{General Coxeter group}
Extension of  commutation relations (\ref{qqc}) to {\it framed Cherednik algebra} is
\be
\label{fqqc}
[q^n_{\ga}\,,q^m_\gb] = -i\epsilon_{\ga\gb}
\left(2 \delta^{nm}I_n+\sum_{v\in \mathcal{R}}\nu(v)\frac{v^{n}v^{m}}{(v,v)}%
\hat K_{v} \right)\,,
\ee
where $I_n$ are idempotents analogous to those of Section \ref{hsmod} and
\be
\hat K_v:= K_v\prod I_{i_1(v)}\ldots I_{i_k(v)}\,,
\ee
where the labels $i_1(v)\,,\ldots \,,i_k (v)$ enumerate those
$I_n$ that carry labels affected by the reflection $\R_{v}$. For instance,
in the case of  $B_p$, $\hat K_{v_{ij}}=K_{v_{ij}}I_i I_j$ for the root $v_{ij}$
generating the permutation of  $e_i$, and $e_j$
 and  $\hat K_{v_i}=K_{v_i} I_i$  for $v_i$ representing the reflection of
 $e_i$ of $B_p$.

It is important that framed Cherednik algebra still possesses inner
$sp(2)$ automorphisms generated by
\be
\label{sp2aut}
t_{\ga\gb}:= \f{i}{4}\sum_{n=1}^p\{q_{\ga}^n\,,q_{\gb}^n\} I_n\,
\ee
obeying (\ref{tqq}). Evidently,
\be
[t_{\ga\gb}\,,I_n]=0\,.
\ee

Note that usual Cherednik algebra results from the framed one by quotioning out the ideal
 identifying all $I_n$ with the unit element of the algebra.

\section{Coxeter higher-spin equations}
\label{CHS}
\subsection{General scheme}

The idea is to let all $x$-dependent
fields $W$, $S$ and $B$ depend on $p$ sets of
variables enumerated by the label $n=1,\ldots p$, that include  $Y_A^n$, $Z_A^n$ ($A=1,\ldots M$),
idempotents $I_n$,
anticommuting differentials $dZ^A_n$
 and Klein-like operators $\hat K_v$ associated with all root vectors
  of a chosen Coxeter group $\C$ (at the convention
 $\hat K_{-v}=\hat K_v$).
 As usual in  HS theory, the fields $W$, $S$ and $B$ are allowed to be valued in
 any associative algebra $A$ (see also Section \ref{Matrix}).  To make contact with the tensorial boundary theory,
it is useful to set $A= (Mat_N)^p$ with elements  represented by
tensorial matrices $a^{u_1\ldots u_p}{}_{v_1\ldots v_p}$, $u_i, v_i = 1\ldots N$.
Here $p$ is the tensor degree of the boundary model while
$N$ is the number of values of color indices with respect to which the $N\to\infty $ limit has to be taken
on the boundary. In this limit the leading contribution should be represented by
the colorless HS theory in the bulk.

The field equations associated with  Cherednik algebra (\ref{qqc}) are
formulated in terms of the star product analogous to (\ref{star2})
\be
\label{star2c}
(f*g)(Z;Y;I)=\frac{1}{(2\pi)^{pM}}
\int d^{pM} S\,d^{pM} T \exp{[iS_n^A T_m^B \delta^{nm} C_{AB}]}\,
f(Z_i+I_i S_i;Y_i+I_i S_i;I)
g(Z-T;Y+T;I)
\ee
with central elements $I_n$ obeying relations
\be
\label{IY}
I_n *Y_A^n= Y_A^n *I_n =Y_A^n\q I_n * Z_A^n=Z_A^n* I_n =Z_A^n\q I_n * I_n =I_n\,.
\ee
With this definition  star product (\ref{star2c}) implies
\be
\label{YZI}
[Y^n_A,Y^m_B]_*=-[Z^n_A,Z^m_B ]_*=2iC_{AB}\delta^{nm} I_n\q
[Y^n_A,Z^m_B]_*=0\,.
\ee
Analogously to Eq.~(\ref{vark}) this star product admits inner Klein operators
$\varkappa_v$ associated with the root vectors $v$
\be
\label{vkv}
\varkappa_v := \exp i\frac{v^n v^m Z_{\ga n} Y^{\ga}{}_m}{(v,v)}
\ee
and any subset of indices $\ga$ among $A$ such that $det |C_{\ga\gb}|\neq 0$.
It is straightforward to see that the Klein operators $\varkappa_v$ generate
the star-product  realization of the Coxeter group  via
\be
\varkappa_{v} * q^n_\ga = \R_{v}{}^n{}_m q^m_\ga * \varkappa_{v}\q q_\ga^n= Y_\ga^n, Z_\ga^n\,.
\ee

Nonlinear equations for the generalized HS theory associated with the Coxeter group $\C$ are
\be
\label{dWc}
\dr W+W*W=0\,,\qquad
\ee
\be
\label{dBc}
\dr B+W*B-B*W=0\,,\qquad
\ee
\be
\label{dSc}
\dr S+W*S+S*W=0\,,
\ee
\be
\label{SBc}
S*B=B*S\,,
\ee
\be
\label{SSc}
S*S= -i \left(dZ^{An} dZ_{An} + \sum_i\sum_{v\in \mathcal{R}_i}F_{i*}(B)
\frac{dZ_n^{\ga}v^{n}dZ_{\ga\,m}v^{m}}{(v,v)}%
*\hat \kappa_v \right)
\,,
\ee
where $\hat \kappa_v$  act trivially on
(\ie commute with) all elements  except for $dZ_{\ga n}$
on which they act in the standard fashion
\be
\hat \kappa_{v} * dZ^n_\ga = \R_{v}{}^n{}_m dZ^m_\ga * \hat\kappa_{v}\,.
\ee
$F_{i*}(B)$ is any star-product function of the zero-form $B$ on the
 conjugacy classes  $R_i$ of $\C$. For instance in the important case of
 $B_p$ equation (\ref{SSc}) reads as
\be
\label{SScb}
S*S= -i \left(dZ_{An} dZ^{An} + \sum_{v\in \mathcal{R}_1}F_{1*}(B)
\frac{dZ_n^{\ga}v^{n}dZ_{\ga\,m}v^{m}}{(v,v)} *\hat\kappa_v
+ \sum_{v\in \mathcal{R}_2}F_{2*}(B)
\frac{dZ_n^{\ga}v^{n}dZ_{\ga\,m}v^{m}}{(v,v)}
*\hat\kappa_v \right)
\ee
with  arbitrary $F_{1*}(B)$ and $F_{2*}(B)$. In particular,
one can set $F_{2*}(B)=0$ keeping $F_{1*}(B)=\eta B$. In the case of $B_1$ this gives
usual HS equations.   $F_{2*}(B)$ can be  nonzero for the $B_p$--HS models
at $p\geq 2$. As discussed in more detail in Section \ref{interpretation}, the roles of
the coupling constants contained in $F_{1*}(B)$ and $F_{2*}(B)$ are different: the former
are responsible for the HS features of the model in $AdS$ while the latter for the stringy
and, more generally, tensorial ones.

Perturbative analysis is  performed around the vacuum solution
\be
\label{vac}
B=0\q S_0 = dZ^{An} Z_{An}\q W=W_0(Y|x)\,,
\ee
where $W_0(Y|x)$ is some solution to (\ref{dWc}) that usually is taken to describe
$AdS$ space-time. (In the $3d$ models of \cite{prok} the solution with
$B=const$ is also important.)

The realization of  $\hat\kappa_{v}$ can be different in different
models and usually needs introduction of   the outer
Klein operators $\hat k_v$  that, by definition, obey (\ref{kq}) with  $q=Y_{\ga n}, Z_{\ga n}$ and
$dZ_{\ga n}$, so that $\hat \kappa_v=\varkappa_v\hat k_{v}$ only affects the differentials $dZ_{\ga n}$.

Equations (\ref{dWc})-(\ref{SSc}) are formally consistent since relations (\ref{qqc})
respect the Jacobi identities which in terms of the field equations are fulfilled
due to the property that the \rhs of (\ref{SSc}) is central.  That HS-type equations can be consistently formulated
based on Coxeter root systems was clear long ago.
(This  was  mentioned already  in \cite{Brink:1993sz}.)
However,  such  equations were never discussed in the literature so
 far since their interpretation  in the standard HS paradigm was not
easy as discussed in Section \ref{hsmod} unless the construction is extended
to the framed Cherednik algebras proposed in this paper.

Multi-particle extensions of the Coxeter HS systems are associated with the semi-simple
Coxeter groups. The simplest option is
associated with the Coxeter root system $B_p^\N$ that consists of the product of
 $\N$  $B_p$ systems
\be
\label{prc}
B_p^\N:= \underbrace{B_p\times B_p\times \ldots}_\N \,.
\ee
The limit $\N\to\infty$ along with the graded symmetrization of the product factors
expressing the spin-statistics relation (for more detail see Section \ref{frames})
is in many respects most natural.

One can consider reductions of the Coxeter HS theories by restricting
all fields to invariants of some group $S$ of automorphisms of $\C$.
A natural choice is some subgroup
$S\subset S_p\times S_\N$ with $S_p \subset B_p$ and $S_\N$  exchanging the product factors in
(\ref{prc}). The latter reduction is possible in the case with all functions $F_i(B)$
associated with different product factors in $\C$ equal to each other.
 Restriction to invariants of $S $ may be important for taking the
$\N\to \infty$ limit.

The choice of $S$ containing the full symmetric group $S_{\N=\infty}$ is distinguished
from various perspectives. If $S=S_\infty$, the resulting algebra is the
(graded symmetric) multi-particle algebra
$M(h(\C))$ of \cite{Vasiliev:2012tv} of the HS algebra $h(\C)$ associated with the
$\C$-framed Cherednik algebra. As explained in \cite{Vasiliev:2012tv}, $M(h(\C))$
is isomorphic to the universal enveloping algebra of $h(\C)$ (see also Section \ref{frames}). This implies in particular
that $M(h(\C))$ is a Hopf algebra. Analogously, if $S=S_\infty\times G$ with $G$ being
some subgroup of the simple Coxeter group $\C$, the resulting multi-particle algebra is
$M(h_G(\C))$ where $h_G(\C)$
is the subgroup of invariants of $G$ in $h(\C)$. Clearly, $M(h_G(\C))$ is also a Hopf algebra.

As discussed in Section \ref{interpretation}, the particularly important case of
 $B_2^\infty$ multi-particle HS model associated with  the
$
sym ( B_2\times B_2\times \ldots )\,
$
system, where the graded symmetrization is  with respect to all elementary  $B_2$ factors,
is anticipated to represent a stringy HS model.

Now we are in a position to
specify different types of the HS models resulting from the proposed construction.

\subsection{Spinor models}
\label{spinmod}

Spinor HS models were formulated in three  \cite{prok} and four
\cite{more} dimensions. Naively, these models may look being too far from Superstring
living in 10 dimensions \cite{Green:1987sp}. This is
indeed true for the original $3d$ and $4d$ HS models  but may not be true
for  their multi-particle extensions. The point is that, as emphasized in
\cite{Gelfond:2003vh,Gelfond:2010pm}, the multi-particle states of a lower-dimensional
model can be identified with elementary states in an appropriate higher-dimensional
theory. Such interpretation turns out to be most natural within the matrix-space approach
to massless field theories elaborated in
\cite{Bandos:1999qf,Vasiliev:2001zy,Vasiliev:2001dc,Bandos:2005mb}
(see also \cite{Vasiliev:2012vf} and recent reviews \cite{Vasiliev:2014vwa,Sorokin:2017irs}).
Moreover, the maximal space-time where such models admit  interpretation in terms
of local fields is ten-dimensional \cite{Bandos:1999qf,Vasiliev:2001dc,Bandos:2005mb}
which is just the superstring space-time dimension. (It would be interesting to see whether
this phenomenon is related to the twistor-like transform in ten dimensions introduced by Witten in
\cite{Witten:1985nt}.)
From this perspective the original
$3d$ and $4d$ theories can be interpreted  as certain branes in the ten-dimensional theory
\cite{Vasiliev:2001dc} with the  $3d$ HS model serving as an elementary
brick from which the others are composed (see also Section \ref{met}).

\subsubsection{$3d$ Coxeter higher-spin models}
The $3d$ Coxeter extension of the models of \cite{prok} is straightforward.
It  still needs
introduction of  two Clifford elements $\psi_{1,2}$ as in \cite{prok}
to induce the doubling of the spectrum of fields as is standard in $3d$
gravity \cite{Achucarro:1987vz,Witten:1988hc} and
HS theory \cite{Blen}. The indices $A$
and $\ga$ coincide, \ie $A$ takes just two values representing a $3d$
spinor index. $sp(2)$ (\ref{sp2aut}) represents the local Lorentz symmetry.
For definiteness we  consider the case of the Coxeter group $B_p$
which seems to be most appropriate for HS applications. Extension to other
Coxeter groups is straightforward.

The multi-particle extension of the $3d$  Coxeter HS theories results from
 endowing all variables with the additional flavour index $a=1,\ldots,\N$
\be
I_n, z_{\ga\,n}, y_{\ga\,n}, dz_{\ga\,n},
\psi_{1,2},\hat \kappa_n, \hat\kappa_{nm}\quad\longrightarrow\quad  I^a_n, z^a_{\ga\,n}, y^a_{\ga\,n}, dz^a_{\ga\,n},
\psi^a_{1,2},\hat\kappa^a_n,\hat \kappa^a_{nm}\,.
\ee
One can either consider different functions $F^a_{1,2*}(B)$ or to keep them
all equal (\ie $a$-independent) to allow the $S_\N$ reduction.
In the latter case  HS equations (\ref{SScb}) read as
\bee
\label{SScbb}
S*S=&& \ls-i\sum_{a=1}^\N \sum^p_{n=1} \Big(dz_{\ga n}^a dz^{\ga n a} + F_{1*}(B)*
{dz_n^{\ga a} dz^a_{\ga\,n}} *\hat\kappa_n\\\nn
 &&\ls+2F_{2*}(B)
*\sum_{m=1}^p\Big (dz^{\ga a}_n (dz^a_{\ga\,n}-dz^a_{\ga\,m}) *\hat\kappa^a_{nm}+
dz^{\ga a}_n (dz^a_{\ga\,n}+dz^a_{\ga\,m}) \hat\kappa^a_n*\hat\kappa^a_{nm}*\hat\kappa^a_n\Big )
\Big)
\,,
\eee
where $\kappa_n$ and $\kappa_{nm}$ are the generators of the Coxeter group $B_p$,
that act  only on $dz^a_{\ga\,n}$.  The resulting
model contains both bosons and fermions for any $\N$ including the $\N=1$ $B_p$--HS
theory. The master fields are demanded to be graded symmetric under the exchange of
odd combinations of spinors $z^a_{\ga\,n}, y^a_{\ga\,n}$ at different $a$ (see also Section
\ref{frames}).

Further reductions can be performed with  $S=S_p$ and/or $S=B_p$. Since
fermions are represented by odd
functions of $y$ the latter choice  leads to a bosonic reduced  model.

In the graded symmetrically reduced model usual HS gauge fields are represented by the $z$-independent
one-form part $\sum^{p,\N}_{i,a}\go (y^a_i, \psi^a_{1i},\hat k_i^a|x)*I_i^a$ of $W$
with $y^a_{\ga i}$ anticommuting at different $a$. In particular, $AdS_3$ background
fields belong to this field
(for more detail see \cite{prok}.) Massless matter fields  are represented by the $z$-independent
part $\sum_{i,a}C (y^a_i, \psi^a_{1i},\hat k_i^a|x)*\psi_{2i}^a$ of the zero-form  $B$. Indeed,
due to the presence of idempotents $I_i^a$ (recall that $\psi_{2i}^a = \psi_{2i}^a* I_i^a$)
these obey the same unfolded field equations as the massless fields of \cite{prok}.
In the absence of  $I_i^a$ the additional terms would contribute changing the pattern of the
equations in agreement with the discussion of Section \ref{hsmod}. In the framed system such terms
are uplifted to the equations on the two-particle fields containing pairs of units $I_{i_1}^{a_1}I_{i_2}^{a_2}$.

As usual in HS theory \cite{more,Vasiliev:1999ba}, one can consider models with fields valued in
matrix algebras \ie carrying Chan-Paton-like matrix indices. Upon imposing appropriate reality
conditions the respective HS models possess $U(n)$ gauge symmetry in the spin-one sector.
Following the construction of \cite{Konstein:1989ij,prok} (see also \cite{Vasiliev:1999ba}) the resulting models
 can be truncated  to those with $O(n)$ and $Sp(2m)$ spin-one gauge groups. (See also Section \ref{Matrix}.)

\subsubsection{$4d$ Coxeter higher-spin models}

In the  Minkowski signature $4d$ Coxeter HS model, $A=\ga,\dga$ and
one introduces two mutually commuting  conjugated algebras generated by  the left
and right  elements $y^n_\ga$, $z^n_\ga$, $\hat k_v$
and  $\bar y^n_\dga$, $\bar z^n_\dga$, $\hat {\bar k}_{\bar v}$, respectively,
on which all fields $W$, $S$ and $B$ depend. In the $4d$ $B_p$--HS system, equation  (\ref{SScb})  is
\bee
S*S= -i \Big(dz_{An} dz^{An} + \ls&&\sum_{v\in \mathcal{R}_1}F_{1*}(B)
\frac{dz_n^{\ga}v^{n}dz_{\ga\,m}v^{m}}{(v,v)} *\hat \kappa_v
+ \sum_{v\in \mathcal{R}_2}F_{2*}(B)
\frac{dz_n^{\ga}v^{n}dz_{\ga\,m}v^{m}}{(v,v)}
*\hat\kappa_v
\nn
\\
\label{SSccb}
&&\ls \ls\ls\ls\ls+ \sum_{\bar v\in \mathcal{R}_1}\bar F_{1*}(B)
\frac{d\bar z_n^{\dga}\bar v^{n}d\bar z_{\dga\,m}\bar v^{m}}{(\bar v,\bar v)} *\hat{\bar \kappa}_{\bar v}
+ \sum_{\bar v\in \mathcal{R}_2}\bar F_{2*}(B)
\frac{d\bar z_n^{\dga}\bar v^{n}d\bar z_{\dga\,m}\bar v^{m}}{(\bar v,\bar v)} *\hat{\bar \kappa}_{\bar v}\Big)\,.
\eee

In fact, there are several  options for the higher-rank extension of the $4d$ HS theory.
The simplest $(B_p\times B_p)'$-option is with identified idempotents $I_n=\bar I_n$ associated with the left and
right oscillators $y^n_\ga$, $z^n_\ga$ and $\bar y^n_\dga$, $\bar z^n_\dga$.
We will call this HS model {\it class $I$}. Recall that by
(\ref{HIK}) both $I_n$ and $\bar I_n$ commute with the Klein operators $\hat k$ and $\bar \hat k$. Hence
$I_n-\bar I_n$ generates an ideal $\I$ of the $B_p\times B_p$ system with independent
$I_n$ and $\bar I_n$. The $(B_p\times B_p)'$ system is $(B_p\times B_p)/\I$.
In this model the lowest states are associated with $4d$ massless fields represented by functions of
a single copy of oscillators $y^n_\ga$, $z^n_\ga,\bar y^n_\dga$, $\bar z^n_\dga$ and $I_n$.
For instance, in the graded symmetrized case the massless fields are described by the $Z$-independent
parts $\go$ and $C$ of  $W$ and $B$, respectively
\be\go=\sum^{p,\N}_{i,a}\go (y^a_i,\hat k_i^a;\bar y^a_i,\hat{\bar k}_i^a|x)*I_i^a \q
\go (y^a_i,\hat k_i^a;\bar y^a_i,\hat{\bar k}_i^a|x) = \go (y^a_i,-\hat k_i^a;\bar y^a_i,-\hat{\bar k}_i^a|x)\,,
\ee
\be
\label{Ck}
 C=\sum^{p,\N}_{i,a}C (y^a_i,\hat k_i^a;\bar y^a_i,\hat{\bar k}_i^a|x)*I_i^a \q
C (y^a_i,\hat k_i^a;\bar y^a_i,\hat{\bar k}_i^a|x) = -C (y^a_i,-\hat k_i^a;\bar y^a_i,-\hat{\bar k}_i^a|x)\,.
\ee

The  $4d$ model with different $I_n^a \neq \bar I_n^a$ which we call {\it class $II$}
  is a particular real form of the complexified $3d$ $(B_p)^2$ system which construction
is fully analogous to the $sl_2(\mathbb{C})$ realization  of the $4d$  Lorentz algebra.
In fact this is a
simplest manifestation of the fact discussed in the beginning of Section \ref{spinmod} that
tensoring of the lower-dimensional models can give higher-dimensional ones. Though this phenomenon
is less straightforward for the higher tensor products since higher-dimensional Lorentz algebras are
not real forms of the direct sums of a number of $sl_2(\mathbb{C})$ the results of
\cite{Bandos:1999qf}-\cite{Sorokin:2017irs} suggest that
a proper interpretation can be available in a certain limit in which
the contribution of (delocalized) branes associated with the lower rank (dimensional)
models can be discarded. In this model, the lower-rank states start with $\go$ and $C$ that depend either only
on $y_n^a,\hat k_n^a$ or only on $\bar y_n^a, \hat{\bar k}_n^a$. These states are analogous to the $3d$ massless
states to be identified with the boundary states in the $AdS_4/CFT_3$ holography once the background
(\ie vacuum) fields are chosen appropriately. This opens an interesting possibility for describing both bulk and
boundary degrees of freedom in the same model. A natural framework for the realization of this
option suggested by the approach of \cite{Vasiliev:2012vf} is described in Section \ref{Proj}.

Finally, there is an interesting further extension of the $4d$ HS system to the $B_{2p}(\mathbb{C})$ -model,
referred to as {\it class III}
which is a specific real form of the complexified $B_{2p}$ $3d$ system where the $\hat K_{i\bar {j}}$ reflection
that exchanges the spinor variables of opposite chiralities is antilinear to be compatible with
the fact that  the dotted and undotted spinors   exchanged
by $\hat K_{i\bar {j}}$ are complex conjugated. Note that this leads to a  parametric freedom
in the model via the phase parameter $\varphi$ in the complex structure  relations
\be
\hat K_{i\bar j} = \exp{i\varphi}\, \hat{\bar K}_{\bar i j}\,.
\ee

For the $4d$ models the natural choice of the symmetric reduction is with $S=S_p$ being the diagonal
subgroup of $S_p\times \bar S_p$ in the left and right sector, allowing
to keep  the frame-like field
\be
\label{frame}
h^{\ga\dga} \sum_{n=1}^p y_{n \ga} \bar y_{n \dga}\,
\ee
 $S$-invariant, as well as its further extension to $S=S_p\times S_\N$ in the
multi-particle case.

\subsection{Vector Coxeter higher-spin models in any dimension}
\label{vect}

Vector HS model in $d$ dimensions of \cite{Vasiliev:2003ev}
is formulated in terms of the oscillators $Z^A_n$, $Y^A_n$ with the double
index $A={\nu,\ga}$ where $\nu= 0,\ldots d$ is the vector index of the
$d$-dimensional $AdS$ algebra $o(d-1,2)$ and $\ga=1,2$ is the vector index
of $sp(2)$  with the symplectic form $\epsilon_{\ga\gb}$ so that
\be
\label{CAB}
C_{AB} = \eta_{\nu\mu} \epsilon_{\ga\gb}\q A=\nu,\ga\q B=\mu,\gb\, ,
\ee
where $\eta_{\nu\mu}$ is the ($x$-independent) $o(d-1,2)$--invariant
fiber metric. The index $\nu$ is decomposed into $\nu=(\nu^\|,\nu^\perp)$
where $\nu^\perp = 0,\ldots,d-1$ labels the Lorentz components  transforming as
a $o(d-1,1)$ vector while  $\nu^\|$, taking one value, labels the Lorentz-invariant
component.
The index $\ga$ of the nontrivial part of Eq.~(\ref{SSc}) is identified with
$\nu=(\nu^\|,\ga)$ (\ie with the $d+1$th component of the $o(d-1,2)$ vector
index $\nu$ in $A=\nu,\ga$) and still takes two values.

The original  HS system of \cite{Vasiliev:2003ev} is formulated in terms
of the star product
\be\label{enlstar} (f *
g)(Z,Y)=\frac{1}{\pi^{2(d+1)}}\int dS dT \, e^{-2S^A_\ga T^\ga_A}
f(Z+S,Y+S)g(Z-T,Y+T)\ ,\ee
which gives
rise to the commutation relations
$$ [Z^\nu_\ga,Z^\mu_\gb]_{*}=-\epsilon_{\ga\gb}\eta^{\nu\mu}\,, \qquad
[Y^\nu_\ga,Y^\mu_\gb]_{*}=\epsilon_{\ga\gb}\eta^{\nu\mu}\,, \qquad
[Y^\nu_\ga,Z^\mu_\gb]_{*}=0\ .$$
The analogue of Eq.~(\ref{SS})  has slightly different normalization
\be\label{vas5} S * S = -\frac{1}{2}(dZ^\ga_A
dZ_\ga^A+4dZ_{\nu\| \ga}dZ^\ga_{\nu^\|}B * \kappa) \,.\ee
$\kappa$ is the product of the inner Klein operator $\varkappa$ and
 the outer Klein operator
$k$ that anticommutes with all $\nu^\|$ components but commutes with the  $\nu^\perp$
Lorentz components.

The nontrivially deformed part of (\ref{vas5})  is of the $B_1$ type.
To extend the $sp(2)$ algebra of automorphisms
of the $B_1$ Cherednik algebra  to the full $sp(2)$ rotations
of $(\nu,\ga)$ with any $\nu$, for $\nu\neq d$
the action of $sp(2)$ is generated by the  $sp(2)$ of
undeformed oscillators. The full $sp(2)$ generators are
\be
t_{\ga\gb}:= -\half\sum_{n=1}^\N\{S_{\nu\ga}^n\,,S_{\mu\gb}^n \eta^{\nu\mu}\}_*\,.
\ee

$t^{tot}_{\ga\gb}$ is the $sp(2)$ generator  acting on all indices
$\ga,\gb$ including those of differentials $dZ_{\nu\ga}$.
In addition, in the model of \cite{Vasiliev:2003ev} it is demanded
that the generator
\be
\label{stot}
t^{int}_{\ga\gb}:=t^{tot}_{\ga\gb} - t_{\ga\gb}
\ee
obeys relations
\be
\dr t^{int}_{\ga\gb}-[t^{int}_{\ga\gb}\,,W]_* =0\q [t^{int}_{\ga\gb}\,,S]_* =0\q [t^{int}_{\ga\gb}\,,B]_* =0\,
\ee
(that restricts the spectrum of dynamical fields
to two-row rectangular Young diagrams of $o(d-1,2)$) and that the fields
in the ideal  formed by the elements of the form $t^{int}_{\ga\gb}*f^{\ga\gb}$
or $f^{\ga\gb}*t^{int}_{\ga\gb}$ have to be factored out,
that restricts the spectrum of dynamical fields
to traceless two-row rectangular Young diagrams of $o(d-1,2)$
appropriate for the description of genuine massless HS fields
in the unfolded formalism \cite{Lopatin:1987hz,Vasiliev:2001wa}.

The Coxeter extension needs several species of oscillators. Here however is a subtlety
that, as  in the simplest $B_1$ case of  \cite{Vasiliev:2003ev},
 the Lorentz components $\nu^\perp$ of the $AdS_d$ vectors should
be treated differently from the   $d+1$th components $\nu^\|$. The latter  are supposed to be affected by the action of the
Coxeter reflections while the former are not to preserve manifest Lorentz covariance.
(Note that the latter condition is  needed in the   Lorentz covariant  setup
but may be relaxed in a more general situation.)
Naively, this suggests that, to use the framed Coxeter algebra
construction, one has to introduce different idempotent elements for the
$d+1$th components of oscillators and the Lorentz ones.
This is however not necessary since Coxeter HS equations
involve hatted Klein operators that obey  (\ref{HIK})
 allowing  to use $o(d-1,2)$-covariant commutation relations
\be
[Y^{\nu\ga}_n\,,Y^{\mu\gb}_m] =  \delta_{nm}\epsilon^{\ga\gb}\eta^{\nu\mu} I_n
\,.
\ee

In terms of these oscillators,
the $o(d-1,2)$ algebra is realized by the
generators
\be
T^{\nu\mu} =-\half\epsilon^{\ga\gb}\sum_{n=1}^\N   Y^{\nu}_{n\,\ga}
Y^{\mu}_{n\, \gb}
\,.
\ee
The $sp(2)$ generators are
\be
t_{\ga\gb}:= -\half  \eta_{\nu\mu} \sum_{n=1}^\N\{S_{n\ga}^\nu\,,S_{n\gb}^\mu \}_* \,.
\ee
The generators $t^{int} $ and $t^{tot}$ are defined as in the original model
using (\ref{stot}).

Analogously to the original HS model, one imposes the $sp(2)$ invariance condition
followed by the factorization condition $t^{int}_{\ga\gb}*f^{\ga\gb}\sim
f^{\ga\gb}*t^{int}_{\ga\gb}\sim 0$. An important remaining question is whether this is
 sufficient to remove all unnecessary states from the general Coxeter HS model making it unitary.
The natural choice of the symmetric reduction is with the symmetric group $S_p$
 that acts on all indices $n$ of $Y_{A n}, Z_{A n}$ and $dZ_{A n}$ irrespectively of the
 value of $A$ and/or $S_\N$ in the multi-particle extension.

Finally let us note that, in agreement with the analysis of
\cite{Vasiliev:2004cm}, it is natural to expect the existence of
another HS model in $d$ dimensions which is holographically dual
to free conformal fermions on the boundary, as well as the further
supersymmetric extension of the bosonic and fermionic models. Though
the explicit form of the nonlinear
field equations was not presented so far, we anticipate that it can be
elaborated along the lines of \cite{Vasiliev:2003ev}. (Recently, some
low-order results in the construction of the fermionic model were presented in \cite{Grigoriev:2018wrx}.)
Such models will also generate a class of Coxeter HS theories in $d$ dimensions.

\section{Higher forms and invariants}
\label{further}

HS theories admit an important extension to higher differential forms along the lines of
\cite{Vasiliev:2015mka}. To this end, it is convenient to unify the $dZ$-- and $dx$--one-forms
 into the master field
\be
\W_1 = \dr_{}+ dx^\un W_\un (Z;Y;\hat K|x)+ dZ^A S_A (Z;Y;\hat K|x)\q \dr_{} =
dx^\un \f{\p}{\p x^\un}\,,
\ee
extending it further to all higher differential forms of odd total degrees with respect to
both $dZ$ and $dx$,
\be
\W = \sum_{p=1,3,\ldots} \W_p\,.
\ee
Analogously, the zero-form master  field $B$ is extended to the field $\B$ containing
differential forms of  all total even degrees
\be
\B =\sum_{p=0,2,\ldots} \B_p\q \B_0 = B\,.
\ee
In these terms the appropriately extended system (\ref{dWc})-(\ref{SSc}) takes
the form
\be
\label{WW}
\W*\W= -i \Big (dZ^{An} dZ_{An} + F_{*}(\B, \gamma_i)
 \Big)\,,
\ee
\be
\label{WB}
[\W\,, \B]_* =0\,,
\ee
where
\be
\label{gi}
\gamma_i =\sum_{v\in \mathcal{R}_i}\frac{dZ_n^{\ga}v^{n}dZ_{\ga\,m}v^{m}}{(v,v)}%
*\hat \kappa_v\,.
\ee
It is important that elements  $\gamma_i$ (\ref{gi}) are central, commuting with
any element of the algebra. Together with (\ref{WB}) this guarantees formal consistency
of the system in the sense that (\ref{WW}) is compatible with
the  associativity of the star product
\be
[\W*\W\,,\W]_*=0\,,
\ee
and the consequence of (\ref{WB})
\be
 [\W*\W\,,\B]_*=0\,.
\ee

In the sector of one-forms $\W_1=W,S$ and zero-forms $\B_0 =B$ this system
with $F_{*}(\B, \gamma_i)$ linear in $\gamma_i$
reproduces (\ref{dWc})-(\ref{SSc}). Higher degrees of two-forms $\gamma_i$
contribute to equations for the higher forms and are important for the construction
of invariants as discussed below.

The construction of equations (\ref{WW}), (\ref{WB})
applies both to irreducible Coxeter systems and to the reducible
ones, containing their direct products. In the latter case the label $i$ of $\gamma_i$
 encodes both the conjugacy classes of the irreducible Coxeter systems and different
factors in the direct product. This corresponds to the extension
$i\to{i,a}$. To impose one or another symmetric reduction,  $F_{*}(\B, \gamma_i)$
should be invariant under the chosen symmetry group $S$.   The appearance of
higher differential forms in the model is important both for its string-like interpretation
since string theory is known to contain higher degree differential forms and for the
construction of invariants along the lines of \cite{Vasiliev:2015mka}.

Invariant functionals are associated with various $x$--space forms $\Ll$ that are closed
as a consequence of field equations (\ref{WW}), (\ref{WB}). As shown in
\cite{Vasiliev:2015mka}, these can be associated with various $dZ$-independent
central elements of the star-product algebra. In the case of usual HS theories
considered in \cite{Vasiliev:2015mka} the only relevant central element was
unity of the star-product algebra. In the Coxeter HS theories
of this paper the $dZ$-independent center of the algebra is generated by the elements
$I_n^a$. In the simplest additive case equation (\ref{WW}) is modified to
 \be
\label{WWL}
\W*\W= -i \big (dZ^{An} dZ_{An} + F_{*}(\B, \gamma_i) \big)
+\Ll
\ee
where
\be
\Ll:= \sum_{n_1,\ldots, n_k,a_1,\ldots, a_k}
I_{n_1}^{a_1}*\ldots * I_{n_k}^{a_k}\, \Ll_{a_1,\ldots, a_k}^{n_1,\ldots, n_k}(x)
 \Big)\,
\ee
obeys
\be
\label{dL}
\dr_{} \Ll_{a_1,\ldots, a_k}^{n_1,\ldots, n_k}(x)=0\,.
\ee
The density forms
$\Ll_{a_1,\ldots, a_k}^{n_1,\ldots, n_k}(x)$ are symmetric under the exchange of
pairs of indices $n_l,a_l$
\be
\Ll_{\ldots, a_n,\ldots, a_m,\ldots }^{\ldots n_n,\ldots, n_m,\ldots}(x)=
\Ll_{\ldots, a_m,\ldots, a_n,\ldots }^{\ldots n_m,\ldots, n_n,\ldots}(x)\,
\ee
and contain differential forms of any positive even degree
\be
\Ll = \Ll_2+\Ll_4+\Ll_6+\ldots
\ee
generating invariant functionals as integrals over space-time cycles of
appropriate dimension
\be
\label{S}
S= \int_{\Sigma}\Ll\,.
\ee

Let us stress that (\ref{dL}) is a consequence of (\ref{WWL}), \ie
apart from imposing field equations on dynamical fields, Eq.~(\ref{WWL}) expresses
$\Ll $ via dynamical fields themselves in such a way that $\Ll$ turns out to be
closed.

 System (\ref{WWL}), (\ref{dL}) and (\ref{WB}) is invariant under the following gauge transformations
with  three types of gauge parameters $\gvep$, $\xi$ and $\chi$ \cite{Vasiliev:2015mka}:
\be
\label{dw}
\delta \W = [\W\,, \gvep]_*\,+  \xi^N \f{\p \F_*(\B\,,\gamma)}{\p \B^N}+
\chi_i \f{\p \Ll}{\p \Ll_i}\,,
\ee
\be
\label{db}
\delta \B = \{\W\,,\xi\}_* + [\B\,, \gvep]_* \,,
\ee
 \be
 \label{dl}
\delta \Ll_i = \dr_{} \chi_i\q
\ee
where $N$ is the  multiindex running over all components of $\B$,
the gauge parameters
$\gvep$ and $\xi$ are differential forms of  even and odd
degrees, respectively, being otherwise  arbitrary functions of $x$
and the generating elements of the star-product algebra, while $\chi_i$ only depend
on $x$ and $dx$. By (\ref{dl}), the functional $S$
(\ref{S}) is gauge invariant.

Generalization of this construction to the systems with non-additive contribution of
the Lagrangian forms $\Ll$ is also possible and can be constructed along the lines of
\cite{Vasiliev:2015mka}.

\section{Color extensions}
\label{Matrix}

As usual in HS theory \cite{more,Vasiliev:1999ba}, one can let all fields be valued in a
matrix algebra, \ie carry Chan-Paton-like \cite{Paton:1969je,Marcus:1986cm}
matrix indices. Upon imposing appropriate reality
conditions the respective HS models possess $U(n)$ gauge symmetry in the spin-one sector.
Following the construction of \cite{Konstein:1989ij,prok} (see also \cite{Vasiliev:1999ba}) the
resulting models can be truncated  to those with $O(n)$ and $Sp(2m)$ spin-one gauge groups.
(For instance  $O(1)$-model is  the so-called minimal HS model that only
 contains even spins.)

 The local symmetry algebras of different types of  $4d$ HS systems of \cite{more}
 are $hu(n,m|8)$, $ho(n,m|8)$ and $husp(n,m|8)$.\footnote{These should not be confused with the
$4d$ global symmetry algebras $hu(n,m|4)$, $ho(n,m|4)$ and $husp(n,m|4)$ of \cite{Konstein:1989ij}.
 The difference expresses the doubling of spinor variables
 $Y\to Y,Z$ in the non-lineal system compared to the $Z$-independent dynamical fields like $\go(Y;K|x)$.}
 Analogously,   local symmetry algebras of different types of  $3d$ HS systems
 are $hu_{\ga\gb}(n,m|4)$, $ho_{\ga\gb}(n,m|4)$ and $husp_{\ga\gb}(n,m|4)$
 ($\ga,\gb =0,1$, $\ga+\gb \neq 0$) \cite{prok}.

 In the both cases the meaning of these notations is that the master fields are valued in appropriate
 $(n+m)\times(n+m)$ matrices in which bosons belong to the diagonal $n\times n$ and $m\times m$
 blocks while fermions are in the two off-diagonal $n\times m$ blocks. The respective HS models
  possess usual supersymmetry at $n=m$.

 For multi-particle HS theories  one can use algebras $M(h...(n,m|2l))$.
 Also one can add matrix indices  to
 $M(h...(n,m|2l))$ in the end which option seems to be less interesting, however.

One  proceeds analogously in the rank-$p$ Coxeter HS models.
The respective algebras will be called  $\C h_{\cdots}(n,m|2l)$ with
$\C$ denoting a  Coxeter group in question. An interesting option
available in this case  is to let the fields $W$ and $B$ be valued in the
matrix algebra $(\times Mat_{N+M})^p$, e.g.,
$C^{u_1\ldots u_p}{}_{v_1\ldots v_p}(Y^n)$, $u_i, v_i = 1\ldots N+M$
 with the idea to assign any pair of matrix indices
$u_n, v_n$ to the variable $Y^n$.
{}From the $AdS/CFT$ perspective, indices $u_i$ with $i=1,\ldots N$ and
$i=N+1,\ldots N+M$ can be associated with the boundary scalar and spinor
fields, respectively. More generally, one can consider multi-indexed fields
with indices $u_i$ and $v_i$ associated with rank--$p$ tensors obeying various
symmetry and/or tracelessness conditions. Still, the respective $\C h_{\cdots}(n,m|2l)$
Coxeter HS theories make sense though the relation of $n$ and $m$ with $N$ and
$M$ is more complicated.

Usually, indices of this type are not considered in the bulk HS
models. The rationale behind this is  that  the singlet
contributions due to the field $C^{u_1\ldots u_p}{}_{v_1\ldots v_p}(Y^n)\sim
\delta^{u_1}_{v_1}\ldots \delta^{u_p}_{v_p} C(Y^n)$ decouple from the traceful
components,  being in a certain sense dominating. This happens because invariants of
the color symmetry cannot source  components  transforming nontrivially
under the color group.
Careful analysis of this phenomenon may be important for understanding specificities
of the $(O(N))^p$-models associated with boundary fields being
symmetric and/or  traceless with respect
to color indices (as well as their unitary and symplectic analogues) in which case the
respective factors $\delta^{u_1}_{v_1}\ldots \delta^{u_p}_{v_p}$ have to be
appropriately symmetrized and/or projected to the traceless components
within the sets of indices ${u_1\ldots u_p}$ and ${v_1\ldots v_p}$.

\section{Idempotent extension}
\label{Proj}

In this section we propose a further extension of the construction of
HS equations allowing to unify  brane-like systems of different
space-time dimensions in the same HS model. This  construction is associated with
idempotents (projectors) of the star-product algebra underlying the system in question.
Usually, these are Fock idempotents. The idea is illustrated by the fact
that the Fock module of the $4d$ HS algebra describes $3d$ conformal fields
\cite{Shaynkman:2001ip} associated with the boundary fields in the $AdS_4/CFT_3$
correspondence \cite{Vasiliev:2012vf}. Let us stress  that in presence of
unbroken HS symmetries, the lower-dimensional branes of this type turn out to be
delocalized, \ie reduction of space-time dimension is via a projective
identification of astigmatic type not allowing to observe certain directions.
However, if the HS symmetries are broken (e.g. by boundary conditions),
the HS branes  can become  space-time localized.

\subsection{General scheme}
The general construction is as follows.  Let $A$ be some associative algebra with the
product $*$ and $\pi_i\in A$
be a set of idempotents
\be
\label{idemp}
\pi_i*\pi_i = \pi_i\,.
\ee
 Let $A_i{}^j\subset A$ be subspaces of $A$ spanned by the elements of the form
 \be
 \label{aij}
a_i{}^j \in A_i{}^j :\quad a_i{}^j =\pi_i *a *\pi_j\,,\quad a\in A\,.
\ee
The  composition law is  of the matrix-like form which, in turn, is a
particular case of this construction
\be
(a*b)_i{}^j = \sum_k a_i{}^k*b_k{}^j\,.
\ee
Note that in this construction the idempotents $\pi_i$ need not
be orthogonal. The only important condition is (\ref{idemp}).

The resulting idempotent-extended algebra $A_{\{\pi\}}$ is associative and unital with the
unit element
\be
\label{IPP}
Id\,{}_i{}^j = \delta_i^j \pi_i\,.
\ee
If the algebra $A$ was unital it can itself be realized as a particular case of the above construction with
$\pi=Id$.

A particular case of this scheme with two idempotents was used in \cite{Vasiliev:1995sv} for the formulation
of a $d=2$ HS theory. Here we propose its slightly  different application directly to the nonlinear
equations of Sections \ref{CHS}, \ref{further}.

\subsection{Realization}

In HS theory, $A$ is  the algebra of functions of
$
dx, dZ, Z, Y, \kappa, x
$
with the star product (\ref{enlstar}) underlying the construction of Section \ref{CHS}.
Idempotents  $\pi_i$ can be identified with the $Z$-independent
Fock projectors of the star-product algebra. We conjecture that exact holographic
correspondence can take place  for the systems which correspond to consistent nonlinear
equations containing  both bulk and boundary fields via the idempotent construction of
this section.

\subsubsection{Vector-like models}

For instance, in the $4d$ HS theory
\be
\label{Fock}
\pi_i = 4I_i \exp  y_{i\ga} \bar y_i^{\ga}
\ee
are such idempotents, \ie
\be
\pi_i *\pi_i = \pi_i
\ee
 (note that here dotted and undotted indices are on equal footing). $\pi_i$ can indeed  be interpreted
 as  Fock idempotents since
 \be
 \label{osc}
 (y_{i\ga} -i\bar y_{i\ga}) * \pi_i =0\q \pi_i* (y_{i\ga} +i\bar y_{i\ga}) =0\,.
 \ee

The idempotent-extended system has the same form as the systems of Sections \ref{CHS}, \ref{further}
with the replacement of the original algebra $A$ by $A_{\{\pi\}}$.
The unit element of the star-product algebra is considered as one of the
idempotents, say, $\pi_0$. Note that with this convention $\pi_0$ is not orthogonal to
$\pi_i$. Luckily, as emphasized in the previous section, they do not need to be orthogonal
for the applicability of the general scheme.
The important consistency condition is that the full set of  idempotents has to
to be invariant under the Coxeter group in the system and the mutual products $\pi_i*\pi_j$
as well as the elements (\ref{aij}) should be well-defined.

In the original HS system, the vacuum solution (\ref{vac}) contains  $S_0$ linear in
$Z$ that solves (\ref{SSc}) because $Z_A$ obey Heisenberg commutation relations (\ref{YZ}). In the
idempotent-extended version of the Coxeter HS equations, the respective vacuum solution
is
\be
S_0 = \sum_i \pi_i * dZ^{An} Z_{An}*\pi_i\,,
\ee
where the condition that all $\pi_i$ are $Z$-independent and hence $*$-commute with $Z_{An}$
implies that so defined $S_0$ solves  (\ref{WWL}) with the unit element (\ref{IPP}).

If HS fields carry matrix indices, the star-product idempotents $\pi^{star}_i$ can
be accompanied by those in color indices $\pi_i^{color}$
\be
\pi_i = \pi_i^{star} \pi_i^{color}\,.
\ee

The simplest idempotent $\pi_i^{color}$  is given by the diagonal
matrix with a single unit element on the diagonal, like $\delta_1^u\delta_v^1$. In that case the left
and right modules in this sector can be associated with the space of rows and columns with respect to
the matrix algebra. This leads to the vector-like fields in $A_0{}^i$ and $A_i{}^0$.

The physical interpretation of the proposed construction is that the fields valued in new idempotent
sectors of  $A_{\{\pi\}}$  live in lower space-time dimensions
compared to those valued in  $A$. This is so since the HS modules supported
by $A_{i}{}^j$-modules are smaller (\ie have smaller Gelfand-Kirillov dimension)
than those of the twisted adjoint $A$-module
supporting dynamical fields in the original system. For instance,
 the $4d$ twisted adjoint module  describes $4d$ massless fields of all spins while
the $A_0{}^i$-module (\ref{aij}), (\ref{Fock})   describes $3d$ conformal fields
(\ie $4d$ singletons) \cite{Shaynkman:2001ip}. This suggests that the idempotent extensions
of the Coxeter HS systems
are appropriate to describe lower-dimensional objects analogous to branes in String Theory.

From the $AdS/CFT$ perspective this construction makes it possible to describe in the same model both
fundamental boundary fields as valued in the   $A_{0}{}^{ i}$ and $A_i{}^0$ sectors and composite current-like
operators as fundamental fields in the twisted adjoint $A$-module. Let us stress again that
the simplest extension of the idempotent construction to the matrix sector gives vector-like boundary fields.

The condition that the set of idempotents $\{\pi_i\}$ is invariant under the action of the Coxeter
group $\C$ restricts significantly the brane (in particular, boundary) sectors.
Since the Coxeter group action exchanges different arguments $Y$ and $Z$ the
condition is that all idempotents $\pi_i^{star}$ that belong to the same orbit of $\C$ should be multiplied
by the same color idempotent, \ie  $\pi_i^{color}= \pi_i^{color}(O_\C (\pi_i^{star}))$, where
$\pi_i^{color}(O_\C (\pi_i^{star}))$ depends on the
 $\C-$orbits of $\pi_i^{star}$-idempotents. For instance, if there is a $\C$-invariant
 $\pi_i^{star}$-idempotent,  it can be multiplied by any $\pi_i^{color}$.
 On the other hand, if there are many different idempotents related by the action of $\C$,
 they should all be multiplied by the same color idempotent.

 The latter situation is  important in the context of usual HS holography.
 In this case idempotents $\pi_i^{star}$ associated with the boundary tensor fields have to
 be of the form
 \be
 \pi^{star}_i= \Pi_i \pi^{color}\,,
 \ee
 where $\Pi_i$ is the Fock projector in the sector of $Y_i$ oscillators and $\pi^{color}$
 has to be the same for any $\pi^{star}_i$. The most natural option is to choose  $\pi^{color}$ as
 the rank-one idempotent
 \be
  \delta_1^{u_1}\delta_{v_1}^1\ldots \delta_1^{u_p}\delta_{v_p}^1\,.
 \ee
Here all indices  $u_1\ldots u_p$ and $v_1\ldots v_p$ have to
take the same number of values $N$. Correspondingly,
the boundary fields are scalars or spinors of the type $\varphi^{u_1\ldots u_p}(x)$ with all indices
taking $N$ values.

\subsubsection{$N=4$ SUSY}

The original and most important example of holographic correspondence
is the duality of Superstring Theory with the most supersymmetric
$N=4$ SYM theory \cite{Maldacena:1997re}. Here we briefly discuss how this
correspondence can be interpreted from the perspective of the proposal of
this paper suggesting that string-like HS theories are associated with the
Coxeter group $B_2$ (see also Section \ref{interpretation}). Remarkably,
the case of $N=4$ supersymmetry turns out to be distinguished from this
perspective as well.

Our consideration is based on the unfolded formulation
of the linearized $N=4$ SYM theory proposed in \cite{Vasiliev:2001zy,Vasiliev:2007yc} where
it was shown that $4d$ conformal massless fields can be described by the
fields valued in certain Fock modules.
Let the Fock vacuum  $\vac_{0\bar0}$ be defined by the relations
\be
\label{xx}
a_\ga * \vac_{0\bar0} =0\,, \qquad \bar b{}^\dgb *\vac_{0\bar0} =0\,,
\qquad \phi_i * \vac_{0\bar0} =0\,,
\ee
\be
\label{xxx}
\vac_{0\bar0} * \bar a^\dga =0 \,,\qquad \vac_{0\bar0} * {b}_\ga =0\,
,\qquad \vac_{0\bar0} * \bar{\phi}^i =0\,
\ee
with respect to the oscillators obeying the following non-zero relations
\be
\label{br}
[ a_\ga ,b^\gb ]_* = \delta_\ga^\gb\,,\qquad
[\bar a{}_{\dot{\gga}} , \bar b{}^\dgb ]_* = \delta_{\dot{\gga}}^\dgb\q
\{\phi_i , \bar{\phi}^j \}_* =\delta_i{}^j\,,
\ee
$i,j=1,\ldots N$. Bilinears of these oscillators form $su(2,2;N)$ which therefore
acts on the Fock module where $4d$ massless fields are valued. In fact, in
\cite{Vasiliev:2007yc} it was shown that, to have a room for HS potentials for
massless fields, one should introduce four Fock modules generated from the Fock
vacua $\pi_{p\bar p}$ ($p,\bar p = 0,1$) obeying relations analogous  to (\ref{xx}) with
exchanged roles of the creation and annihilation operators $a_\ga$, $b_\gb$ and
$\bar a^\dga$, $\bar b^\dgb$. For instance,
\be
\label{txx}
b_\ga * \vac_{1\bar 1} =0\,, \qquad \bar a{}^\dgb *\vac_{1\bar 1} =0\,,
\qquad \bar\phi^i * \vac_{1\bar 1} =0\,,
\ee
\be
\label{txxx}
\vac_{1\bar 1} * \bar b^\dga =0 \,,\qquad \vac_{1\bar1} * {a}_\ga =0\,
,\qquad \vac_{1\bar 1} * {\phi}_i =0\,.
\ee

The Clifford oscillators can be interpreted as
generating the color matrix algebra $Mat_{2^{2N}}$ where all fields are valued.
Let us now explain how this construction appears in the framework of the $B_2$-HS model
and why the case of $N=4$ is distinguished.

The $B_2$-HS theory contains a pair of oscillators $y_{i\ga}$, $\bar y_{i\dga}$.
This allows us to build oscillators analogous to (\ref{osc})
\be
a_\ga =  y_{1\ga} +i y_{2\ga}\q b_\ga = \f{1}{4i} ( y_{1\ga} -i y_{2\ga})\q
\bar a_\dga =  \bar y_{1\dga} -i \bar y_{2\dga}\q
\bar  b_\dga = \frac{1}{4i} (\bar y_{1\dga} +i\bar  y_{2\dga})
\ee
defining vacuum $\vac_{00}$ (\ref{xx}). More precisely, since the commutation relations
(\ref{YZI}) contain  $I_n$, to obey (\ref{xx}),  (\ref{xxx}) and (\ref{idemp})
the idempotent $\pi_{00}$ should contain the factor of $I_1*I_2$.

The fields valued in the left and right modules
generated from this vacuum  describe $4d$ massless conformal fields. To describe
supermultiplets the vacuum $\vac_{0\bar0}$ has to be combined with the Clifford vacuum also
obeying (\ref{xx}), (\ref{xxx}) \cite{Vasiliev:2001zy}.

Although, naively, this construction relates the $B_2$--HS theory to the massless
conformal supermultiplets at any $N$, this is not quite the case. The problem  is
that the idempotent $\vac_{0\bar0}$ defined this way is not invariant under the $B_2$ reflection
exchanging $Y_{1}^A$ with $Y_2^A$ or changing a sign of one of them. Indeed, such
reflections map  idempotent $\vac_{0\bar 0}$ to the {\it opposite idempotent} $ \vac_{1\bar1}$.
In fact, the condition that the set of idempotents has to be invariant under the action
of the Coxeter group $B_2$ demands all $\vac_{p\bar p}$ be present
in the model.

The problem however is that opposite idempotents  may have ill-defined mutual star product
so that elements $\vac_{0\bar0}*a* \vac_{1\bar1}$ (\ref{aij}) are ill defined.
Namely, for purely bosonic opposite idempotents of this type, \ie at $N=0$,
 $\vac_{0\bar0}* \vac_{1\bar1}$ is divergent. On the other hand, in the purely Clifford
 case the product of  opposite Fock idempotents is zero. In the supersymmetric case
 the bosonic and fermionic contributions to the vanishing determinants in the
 respective Gaussian integrals resulting from the star product of the idempotents
  have opposite signs. The full compensation of the
 bosonic divergency occurs just at $N=4$ when the numbers of bosonic and fermionic
 oscillators are equal.  (Note that for a similar reason, at $N=4$ the conformal SUSY is
  $psu(2,2;4)$.)

Thus the consistent rank-two conformal boundary
system in the $B_2$--HS theory  must have $N\geq 4$.
The modules generated from $\vac_{p\bar p}$
 describe $4d$ conformal massless fields (supermultiplets) of all spins
 \cite{Vasiliev:2001zy,Vasiliev:2007yc}. It is natural to
conjecture that all of them except for those in the lowest-spin $N=4$ SYM supermultiplet
will become massive upon spontaneous breakdown
of HS symmetries in the multi-particle $B_2$--HS theory.
Note that simultaneously the non-zero string tension should result from the spontaneous
  breaking of HS symmetries

At the
condition that the resulting theory is free from  massless fields of
spins $s\geq 2$, the only remaining option is the $N=4$ SYM multiplet.
  This is the case of $N=4$ SYM which is the
 only $N=4$ massless conformal system with spins $s\leq 1$. Thus, a spontaneously
 broken $B_2$-HS theory with non-zero string tension has a chance to be dual to $N=4$ SYM.

 To have a nontrivial gauge group for the $N=4$ SYM boundary theory of the spontaneously
 broken HS theory one has to
 introduce color indices additional to the Clifford algebra. This can  be achieved
 either directly by letting all fields be valued in the matrix algebra (and its
 further orthogonal/symplectic reductions) or
 via the construction used for the vector models introducing the fields
 $C^{u_1 u_2}{}_{v_1 v_2}(Y)$ and using the additional color idempotent
 \be
 \pi^{col}= \delta^{u_1 u_2}\delta_{v_1 v_2}\delta^{u_1}_{v_1}\,.
 \ee

\section{Unitarity}
\label{unitarity}

One of the advantages of the unfolded formulation of dynamical equations
is that it makes symmetries manifest operating directly with modules of the
symmetry $h$ underlying the model in question. In particular, linearized equations
on the zero-forms in the system have a form of covariant constancy conditions
\be
D_0 C^I(x) =0
\ee
with $C^I(x)$ valued in some  $h$-module $V$, and $D_0$ being
 a flat covariant derivative of $h$. Though  $V$ is not unitary in the
 Lorentz-covariant frame, the condition that the system admits  consistent quantization
 compatible with unitarity demands $V$ be complex equivalent to some unitary $h$-module
 $U$ \cite{Vasiliev:2001zy}. This makes it straightforward to analyze the pattern of one
 or another unfolded system in terms  of space-time symmetry algebra modules  including
 the issue of unitarity.

Naively, the proposed Coxeter HS systems
contain non-unitary sectors associated with the tensor products of the
unitary HS modules with the so-called topological modules of the original
HS theory, that describe non-unitary finite-dimensional  modules of the $AdS$
algebra.

For instance, in the $4d$ $B_2$-HS theory, consider a rank-two field $C(Y_1,k_1;Y_2, k_2) := C_{0,1} (Y_1,Y_2)  k_2$
 proportional to the Klein operator $k_2$ of the second type and independent of $k_1$.
The covariant derivative with respect to  $AdS$ background connection
contains commutator with the frame field $h(Y)$ with respect to the first argument
and anticommutator with respect to the second.
This is manifested in the form of covariant derivative
\be
\label{d0}
D_0 (C_{1,2}(Y_1,Y_2)) =
 D^\text{L}C_{1,2}(Y_1,Y_2)   - {h}^{\ga\dgb}\left(y_{1\ga} \frac{\partial}{\partial \bar{y}_1^\dgb}
+ \frac{\partial}{\partial {y}_1^\ga}\bar{y}_{1\dgb} -
i y_{2\ga} \bar{y}_{2\dgb} +i \frac{\partial^2}{\partial y_2^\ga
\partial \bar{y}_2^\dgb}\right)C_{1,2}(Y_1,Y_2) \,
\ee
with the Lorentz covariant derivative
\be
D^\text{L} A  = \dr_{}  -\sum_{i=1,2}
\left(\go^{\ga\gb}y_{i\ga} \frac{\partial}{\partial {y}_i^\gb} +
\bar{\go}^{\dga\dgb}\bar{y}_{i\dga} \frac{\partial}{\partial \bar{y}_i^\dgb} \right)\,.
\ee
The first term in (\ref{d0}) acts  on homogeneous polynomials of $Y_1$ of any definite degree while the
second mixes polynomials of different degrees of $Y_2$. This structure is tantamount to the fact that
 $C_{1,2}(Y_1,Y_2)$ is valued in the tensor product of the adjoint module of the HS algebra
 with respect to $Y_1$ and twisted adjoint module with respect to $Y_2$. The twisted adjoint module
 and its tensor products  correspond to unitary particle-like states \cite{Ann} and their
 multi-particle tensor products, respectively.

 Having a form of the infinite sum of finite-dimensional modules, the zero-form fields valued in the
 adjoint module or even containing it as a factor as in (\ref{d0}) cannot correspond to a
 unitary particle-like representation.  The only exception is when the field
 $C_{1,2}(Y_1,Y_2)$ is independent of
 $Y_1$. Then it contains the factor of $I_1$ and,  in accordance with the
 discussion of Section \ref{CHS}, corresponds to unitary  massless states of
 the usual HS theory in the sector of variables $Y_2$.

 Naively, non-singlet states in the adjoint module factors   break down unitarity
 of the system. However,  these potentially non-unitary states can
 be consistently truncated away. Indeed, consider for instance product of
 the factors $C_{1,2}(Y_1,Y_2) k_2$. There are two options. Either the arguments $Y_1,k_1$ and
 $Y_2, k_2$ are permuted, in which case one ends up with the unitary rank-two field, or they
 are not permuted. In the latter case, the product will still contain the singlet component $I_1$
 in the adjoint factor describing nonlinear corrections to usual  massless HS equations,
 \ie non-singlet elements of the adjoint factor are never generated. Note that, to
 account properly the contribution of the Klein operators, their explicit appearance
 in Eq.~(\ref{WW}) via the factors of $\gamma_i$ should be taken into account.

In the original HS theory topological fields were interpreted as
moduli (coupling constants) rather than propagating degrees of freedom.
Similarly, in the Coxeter HS theories,  the fields
associated with non-unitary sectors resulting from the tensor products of
unitary and topological modules should be interpreted as non-propagating
by imposing appropriate boundary conditions in the topological sector
somewhat analogously to the analysis of conformal gravity in \cite{Maldacena:2011mk}.
 On the other hand assigning non-zero VEVs to the fields in potentially
non-unitary sectors may significantly affect the interpretation of the bulk HS theory
similarly to the $3d$ HS model of \cite{prok} in which the value of the singlet
topological field is interpreted as the mass parameter  in the theory.
Analogous suggestion for the mass-generating spontaneous breaking mechanism in
multi-particle HS theories is discussed in Section \ref{interpretation}.
The  question  whether the non-unitary fluctuations can be consistently
truncated away in presence of non-zero VEVs of topological fields remains to be investigated.

A related comment is that the issue of whether a HS model is unitary or not may  depend on
the choice of reality conditions and vacuum solution. For instance, in the sector of zero-forms $C$
of the $4d$ HS theory,  condition (\ref{Ck}) just implies that, from the perspective of the $3d$ model,
the field $C$  is valued in the tensor product of (complexified) $3d$ particle module with the topological
one. Nonetheless, the resulting module of the $4d$ HS model turns out to be
unitary due to imposing appropriate reality conditions compatible with the $4d$ vacuum solution.

To summarize, to make the proposed models unitary,  potentially dangerous
fields containing products with non-singlet factors of the adjoint HS module
have to be truncated away. Remarkably, this
is possible in all orders so that the reduced model turns out to be unitary.
This factorization is somewhat analogous to  that of non-singlet color sectors
in the HS theory with colorful fields.

Let us stress that the HS models dual to
the free tensor boundary theories with $O(N^p)$ broken to $O(N)^3$ at the level of
observables have to be unitary \cite{Beccaria:2017aqc}. This condition has little to do with potential
non-unitarity of interacting tensor models with ill-defined potentials. On the other hand,
it would be interesting to see whether the non-unitary boundary tensor models can be dual
to $B_p$-HS theories with  non-unitary bulk fields switched on.

\section{Field frames of  multi-particle theories}
\label{frames}
In this section we recall the construction of  \cite{Vasiliev:2012tv} for the multi-particle algebra
 extending it to the graded-symmetric case and analysis of  idempotents.

\subsection{Definition and field equations}
\label{def}
Let $A$ be an associative algebra with the product law $*$ and basis elements $t_i$ obeying
\be
\label{struc}
t_I * t_J = f^K_{IJ} t_K\q
f_{IJ}^K f_{KL}^N = f_{IK}^N f_{JL}^K\,.
\ee
In the Coxeter HS case, $A$ is  (matrix-valued extension of) the algebra of functions
of $I_i, Z^A_i, Y_i^A, K_v, dZ^A_i$
as well as of $\pi_i$ for the idempotent extension of Section \ref{Proj}
(space-time differential $ dx^\un$ is not included).

As a linear space, multi-particle algebra $M(A)$ is  the direct sum
of all graded-symmetric tensor degrees of $A$
\be
\label{tens}
M(A) = \sum_{n=0}^\infty \oplus Sym \,  \underbrace{A\otimes \ldots \otimes A}_n\,,
\ee
where the $\mathbb{Z}_2$-grading is identified with the power in $Z^A_i$ and $Y_i^A$,
\be
f(-Z^A_i, -Y^A_i)= (-1)^{\pi_f}f(Z^A_i, Y^A_i)\,,
\ee
\ie odd polynomials in $Z^A_i$ and $Y_i^A$ are antisymmetrized under the symbol $Sym$
in (\ref{tens}). Let us stress that this does not contradict to the fact that
$Z^A_i$ and $Y_i^A$ obey star-product commutation relations in $A$.
So defined graded symmetrization accounting the total number of spinorial indices
carried by $Z^A_i$ and $Y_i^A$  expresses the Pauli principle. (Note that in the vectorial
model of \cite{Vasiliev:2003ev} considered in Section \ref{vect} only integer-spin states
contribute.)

Let $\ga_I$ be odd (even) elements of the Grassmann algebra associated with the
odd (even) elements $t_I$, respectively,
\be
\label{grcom}
\ga_I \cdot \ga_J = (-1)^{\pi_I \pi_J}\ga_J\cdot\ga_I\,.
\ee
The product $\cdot$ denotes usual (juxtaposition) product in the Grassmann algebra
of $\ga$   introduced  to distinguish between the graded symmetric
tensor product  $\cdot$ and the commutative product in the definition of the
star-product algebra  underlying $A$ of $M(A)$.
The latter is the algebra of functions $F_\cdot(\ga)$
with the product law
\be
\label{circ}
F_\cdot(\ga)\circ G_\cdot(\ga) = F_\cdot(\ga)\cdot  \exp_\cdot \Big (  \f{\overleftarrow{\p}}{\p \ga_I} f^N_{IJ} \ga_N
 \f{\overrightarrow{\p}}{\p \ga_J} \Big )\cdot G_\cdot(\ga)\,.
\ee
An elementary computation shows  that associativity of $A$ implies associativity of
the product $\circ$ of $M(A)$. (For the even case see \cite{Vasiliev:2012tv}.)

Algebra $M(A)$ is unital, with the unit element $Id$ identified with $F(\ga)=1$.
The $\mathbb{Z}_2$--grading of $M(A)$ is induced by that of $A$
\be
F((-1)^{\pi (\ga)}\ga)= (-1)^{\pi(F)} F(\ga)\,.
\ee
The graded commutativity (\ref{grcom}) implies the graded symmetrization in (\ref{tens}).

The power-series expansion of the functions $F_\cdot(\ga)$
\be
F_\cdot(\ga)=\sum_{n=0}^\infty F^{I_1\ldots I_n}\cdot \ga_{I_1} \cdot \ldots \cdot \ga_{I_n}
\ee
will be assumed to have anticommuting coefficients $F^{I_1\ldots I_n}$
\be
\label{pauli}
F^{I_1\ldots I_n}\cdot G^{J_1\ldots J_m} = (-1)^{(\sum_{k=1}^n \pi_{I_k})(\sum_{l=1}^m\pi_{J_l})}
G^{J_1\ldots J_m} \cdot F^{I_1\ldots I_n}\,.
\ee
The same time it is convenient to demand them to commute with $\ga_I$
\be
F^{I_1\ldots I_n}\cdot  \ga_J = \ga_J \cdot F^{I_1\ldots I_n} \,.
\ee
Since  components of  $F_\cdot (\ga)$ are identified with the physical
fields in the multi-particle HS theory,  convention (\ref{pauli}) expresses the  Pauli principle.

Since $A\subset M(A)$ is represented by linear functions of $\ga_I$, the  $*$
product acts on linear functions of $\ga_I$ according to
\be\label{staralfa}
\ga_J *\ga_K  =  f_{JK}^M\ga_M \,.
\ee
Note  that $A\circ A$ does not belong to $ A$.

 $M(A)$ is isomorphic to the universal enveloping algebra $U(l(A))$,
\be
\label{iso}
M(A)\sim U(l(A))\,,
\ee
where  $l(A)$ is the Lie superalgebra associated with $A$,
\be
[t_I\,,t_J]_* = g_{IJ}^K t_K\q
g_{IJ}^K = f_{IJ}^K- (-1)^{\pi_I\pi_J}f_{JI}^K\,.
\ee
This is because  $M(A)$
is isomorphic to $U(l(A))$ as a linear space and
\be
\label{tcirct}
\ga_I\circ \ga_J - (-1)^{\pi_I\pi_J}\ga_J\circ \ga_I = g_{IJ}^K \ga_K\,.
\ee
Concise form of the product law (\ref{circ}) is specific for the case of
 Lie superalgebras  $l=l(A)$ associated with
associative algebras $A$.

\subsection{Frame generating functions}
\label{framegen}
 The following useful property of $M(A)$  is a simple
consequence of Eq.~(\ref{circ})
\be
\label{fr} \forall f,g\in
A:\qquad \exp_\cdot f(\ga) \circ \exp_\cdot g(\ga) =
\exp_\cdot(f\bullet g)(\ga)\,,
\ee
where
\be \label{bull} f\bullet g := f+g +f* g =
(f+e_*)* (g+e_*)-e_*\in A\,
\ee
and $e_*$ is the unit element of $A$ if the latter is unital
 (recall that $f,g\in A$ implies that $f(\ga)$ and $g(\ga)$ are linear
in $\ga$).
Associativity of $*$ implies associativity of $\bullet$
 even if $A$ is not unital.

Let
\be
\label{fnu}
G_{\cdot \nu} =\exp_\cdot (\nu)\in M(A)\q \nu=\nu^I\cdot \ga_I\,,
\ee
where $\nu^I$ are free Grassmann-odd (even) parameters
for odd (even) $\ga_I$, which however commute with $\ga_i$
\be
\nu^I\cdot \nu^J =(-1)^{\pi_I\pi_j} \nu^J\cdot \nu^I\q \nu^I\cdot \ga_J
=\ga_J \cdot \nu^I\,.
\ee

Eq.(\ref{fr}) gives
\be
\label{fnumu}
G_{\cdot \nu}\circ G_{\cdot \mu} = G_{\cdot \nu\bullet \mu}\,.
\ee
This formula is convenient for practical computations via differentiation
over $\nu^I$ with $G_{\cdot \nu}$  serving as
the generating function for elements of $M(A)$.

As any universal enveloping algebra $M(A)$ is double filtered. Indeed,
let $V_n$ be the linear space of order-$n$  polynomials of $\ga_I$.
{}From Eqs.~(\ref{circ}), (\ref{tcirct}) it follows that
\be
F_n \circ F_m \in V_{n+m}\q F_n \circ F_m -(-1)^{\pi(F_n)\pi(F_m)}F_m \circ F_n \in V_{n+m-1}
\q F_{m}\in V_{m}\,,\quad F_{n}\in V_{n}\,.
\ee

A  linear map of $M(A)$ to itself can be represented by
\be
\U(\ga, a) [F_\cdot]= \U(\ga, \f{\overrightarrow{\p}}{\p \ga}) F_\cdot(\ga)\,,
\ee
where derivatives $\f{\overrightarrow{\p}}{\p \ga}$ act on $F_\cdot(\ga)$.
This formula can be interpreted as representing the action of the normal-ordered
oscillator algebra with the generating elements $\ga_I$ and $\bar\ga^J$
acting on the Fock module spanned by $F(\ga)|0\rangle$ with
$\bar\ga^I|0\rangle =0$.
To respect the double filtration property mapping order-$n$ polynomials to
order-$n$ polynomials, $\U(\ga, \bar\ga)$ should be {\it filtered} obeying
\be
\U(\ga, \bar\ga)=\sum_{m,n=0}^\infty \U^{I_1\ldots I_m}{}_{J_1\ldots J_n}\ga_{I_1}\cdot \ldots
\cdot\ga_{I_m}\cdot \bar\ga^{J_1}\cdot \ldots\cdot  \bar\ga^{J_n}\,
\q
\U^{I_1\ldots I_m}{}_{J_1\ldots J_n}=0 \quad \mbox{at}\quad m>n\,.
\ee

In these terms, the unit map is
$
\mathbf{Id} = 1\,.
$
Consider maps of the form
\be
\label{Uf}
U(f) \equiv \U(\ga,\bar\ga|f)= \phi \exp_\cdot (\ga_I\cdot  f_\cdot^I(\bar\ga))
\ee
with some $\ga$-independent coefficients $f_\cdot^I(\bar\ga)$ and constant $\phi$.
The map $U(f_\cdot)$ is filtered if $f_\cdot^I(\bar\ga)$ is at least linear in
$\bar\ga$, \ie
\be
\label{atl}
f_\cdot^I(0) =0\,.
\ee

Interpreting
$\bar\ga$ as parameters, we can identify any $f(\ga):=\sum_I f^I \ga_I $
with $f(t)=\sum_I f^I t_I \in A$. For $U(f)$ (\ref{Uf}) acting on $G_{\cdot\nu}$ (\ref{fnu})
we obtain
\be
\label{uf}
U(f)(G_{\cdot\nu}) =\exp_\cdot (\tilde f^I(\nu)\ga_I)\q \ \tilde f^I(\nu)=
\nu^I +f^I(\nu)\,.
\ee
Hence, Eq.~(\ref{fr}) gives
\be
\label{ufuf}
U(f)(G_{\cdot \nu})\circ U(g)(G_{\cdot\mu}) =
\exp_\cdot((\tilde f(\nu)\bullet \tilde g(\mu))(\ga))\,,
\ee
where $\tilde f(\nu)$ and $\tilde g(\mu)$ are interpreted as elements of $A$,
\ie
$
\tilde f(\nu) =
\tilde f^I(\nu)\ga_I\,.
$

Consider map (\ref{Uf}) with
\be
f^I(\bar\ga)\ga_I = f(a_n)\q a_{n} :=
\ga_{I_1}*\ldots *\ga_{I_n}\cdot \bar\ga^{I_1}\cdot \ldots\cdot  \bar\ga^{I_n}\q a_{1}=a=\ga_I \cdot \bar\ga^I\q a_0=e_*  \,,
\ee
where $f(a_n)$ is a linear function of $a_n$ ($n\geq 1$).
Such a map has the form (\ref{Uf}) since $a_n\in A$.
A particularly important class of maps (\ref{Uf}) is represented by
$\mathbf{U}_{u}$ of the form
\be
\label{ubgb}
\mathbf{U}_{u}(a) = :\exp[ u(a) -a]:\,,
\ee
where normal ordering is with respect to $\ga_I$ and $\bar\ga^I$ and for $a\in A$
\be
\label{u(f}
u(a)=
(u_1{}^1 a+u_1{}^2 e_* )*
(u_2{}^1 a+ u_2{}^2 e_*)_*^{-1}\q (e_* +\gb a)_*^{-1} :=
\sum_{n=0}^\infty (-\gb
)^n a_n \,
\ee
with $u_i{}^j\in \mathbb{R}$. Composition of such maps gives  a map of the
same class
\be
\mathbf{U}_{u}\mathbf{U}_{v}= \mathbf{U}_{uv}\,,
\ee
where $(uv)_i{}^j= u_i{}^k v_k{}^j$ is the matrix product.
The maps $\mathbf{U}_{u}$ with $det |u|\neq 0$ are invertible and form
  Mobius group.

{}From Eq.~(\ref{uf}) it follows that
\be \label{munu} \mathbf{U}_{u}
(G_{\cdot \nu}) = G_{\cdot {u}(\nu)}\,.
\ee
To find the composition law of $M(A)$ in the frame $T_{I_1\ldots I_n}^{u}$
associated with
$
G_{\cdot{u}(\nu)}
$
via
\be
\label{bas}
T_{I_1\ldots I_n}^{u} = \f{\p^n}{\p \nu^{I_1}\ldots \p \nu^{I_n}}
G_{\cdot{u}(\nu)}\Big |_{\nu=0}\,
\ee
we compute
\be
G_{\cdot\nu} \diamond G_{\cdot\mu} :=
\mathbf{U}^{-1}_{u} ( G_{\cdot{u}(\nu)} \circ G_{\cdot{u}(\mu)})\,.
\ee
Eq.~(\ref{bull}) gives
\be
\label{diam}
G_{\cdot\nu} \diamond G_{\cdot\mu} =G_{\cdot u^{-1}(u(\nu)\bullet u(\mu))} \,.
\ee

General map (\ref{ubgb}) is not filtered, not respecting  condition
(\ref{atl}). The subgroup $P$ of filtered maps (\ref{ubgb}) is represented by the
lower triangular matrices
\be
\label{aff}
u_{b,\gb}(f) = b f* (e_*+\gb f)^{-1}_*\,
\ee
with the composition law
\be \label{grlow} b_{1,2}=b_1 b_2\q \gb_{1,2} = \gb_2 +\gb_1 b_2\,.
\ee

Using for these transformations  notation
$U_{b,\gb}$ instead of $U_u$ we observe that the unit element is
\be
\mathbf{Id} =\mathbf{U}_{1,0}\,
\ee
and
\be
\label{inv}
\mathbf{U}^{-1}_{b,\gb}= \mathbf{U}_{b^{-1}, -\gb b^{-1}}\,.
\ee

The map
\be
\label{RU}
{\mathbf{R}} =\mathbf{U}_{-1,1}\,
\ee
 is involutive
\be
\label{mathr}
\mathbf{R}^2 = \mathbf{Id}\,
\ee
describing the principal  antiautomorphism of $M(A)$  \cite{Vasiliev:2012tv}.

For the maps (\ref{aff}), the composition law (\ref{diam}) takes the form \cite{Vasiliev:2012tv}
\be
\label{diamaf}
G_{\cdot \nu} \diamond G_{\cdot \mu} =G_{\cdot\gs_{b,\gb}(\nu,\mu)} \,,
\ee
where
\be
\label{sig}
\gs_{b,\gb}(\nu,\mu) = -\gb^{-1} (e_*-(e_*+\gb \mu )* (e_*-\gb (b+\beta) \nu* \mu )^{-1}*
(e_*+\gb \nu ))\,.
\ee
Three most important cases include
\be
\label{10}
\gs_{1,0}(\nu,\mu) = \nu+\mu +\nu* \mu=\nu\bullet \mu\,,
\ee
\be
\label{01}
\gs_{-1,1}(\nu,\mu) = \nu+\mu +\mu * \nu= \mu\bullet \nu\,,
\ee
\be
\label{fus}
\gs_{1,-\half}(\nu,\mu) = 2(e_*-(2e_* -\mu)* ( 4e_*+ \nu* \mu )_*^{-1}
* (2e_*-\nu))\,.
\ee

Here $\gs_{1,0}(\nu,\mu)$ corresponds to the identity map
reproducing the  basis in $M(A)$ resulting just from the
 symmetrized tensor product of the framed oscillator algebras. $\gs_{-1,1}(\nu,\mu)$
corresponds to the basis resulting from the action of principal
 antiautomorphism $\mathbf{R}$.   The case of  $\gs_{1,-\half}(\nu,\mu)$
is most interesting, reproducing the current operator algebra of \cite{Gelfond:2013xt}.

Practically, the difference between the frames (\ref{10}) and (\ref{fus}) is as
follows.
The product law associated with (\ref{10}) is the original product law
(\ref{grcom}), from which it is easy to see that the product $\circ$ of two
polynomials of $\ga$ of degrees $n$ and $m$ gives degree-$k$ polynomials with
$max(n,m)\leq k\leq n+m$. This implies that, with the product law
(\ref{grcom}), the higher-rank fields (say, rank-two)
will not contribute to the equations for the lower-rank ones (say, rank-one).
This property is inconsistent both with structure of the boundary operator
algebra analyzed in \cite{Gelfond:2013xt} and with the idea that VEVs of the
higher-rank fields can modify field equation for the lower-rank ones to yield
a Higgs phenomenon. The product law associated with (\ref{01}) has analogous
properties.

However, in \cite{Vasiliev:2012tv} it was shown that the multi-particle algebra
in the frame (\ref{fus}) properly reproduces the boundary operator algebra inducing
such a product  law that the product  of two
polynomials of $\ga$ of degrees $n$ and $m$ gives degree-$k$ polynomials with
$|m-n|\leq k\leq n+m$ which  is consistent both with the structure of the
boundary operator algebra of  \cite{Gelfond:2013xt} and  with the idea of higgsing
by virtue of VEVs of  higher-rank fields. Hence, we anticipate that frame (\ref{fus})
 is most appropriate for the analysis of the multi-particle Coxeter HS theories.
Let us stress that since the multi-particle
algebra is infinite dimensional it is not {\it a priori} guaranteed that the multi-particle
HS theories associated with different frames described in this section are physically
equivalent.

In application to nonlinear equations of the multi-particle HS theory,  system (\ref{WW}), (\ref{WB})
preserves its  form with the $*$  replaced by $\diamond$
\be
\label{WWd}
\W\diamond\W= -i \Big (dZ^{An} dZ_{An} + F_{\diamond}(\B, \gamma_i)
 \Big)\q
[\W\,, \B]_\diamond =0\,,
\ee
where $\diamond$ is built via (\ref{fus}) from the star product of the Coxeter HS theory
underlying its multi-particle extension.

\subsection{Unity idempotent map}
\label{uid}

Multi-particle algebra has an important property that every idempotent $\pi \in A$
induces an idempotent $\Pi\in M(A)$. This immediately follows from formulae (\ref{fr}),
(\ref{bull}) with
\be
\Pi := \exp_\cdot -\pi\,,
\ee
\ie $\Pi\circ \Pi = \Pi$ if $\pi*\pi = \pi$. In particular, application of this
construction to $\pi = e$ gives the {\it unity idempotent} of $M(A)$
\be
\label{pie}
\Pi_e:= \exp_\cdot -e\,.
\ee
Note that being built from the central element $e$, $\Pi_e$ is central in $M(A)$.

Unity idempotent $\Pi_e$ has an interesting property that it is
$\circ$ -orthogonal to any $a\in A$
\be
\Pi_e\circ a = a\circ \Pi_e = 0\q a\in A\,,
\ee
which is obvious from (\ref{circ}) and (\ref{pie}).
Analogously one can check that
\be
\label{apib}
(a\cdot \Pi_e) \circ b= (a*b)\cdot \Pi_e\q \forall a\in A
\ee
and
\be
\label{apibpi}
a\cdot \Pi_e \circ b \cdot \Pi_e = a*b \cdot \Pi_e\q \forall a,b \in A\,.
\ee
This relation provides a homomorphism of $A$ to $M(A)$. It should be stressed
however that this map is not polynomial. (No homomorphism of $A$ to $M(A)$ realized as
polynomial functions of $A$ exists.)

Relation (\ref{apibpi}) has the consequence that
\be
\Pi_e^1 := e\cdot \Pi_e
\ee
is also an idempotent
\be
\Pi_e^1 \circ \Pi_e^1 =\Pi_e^1\,.
\ee
From (\ref{apib}) it follows
\be
\Pi_e^1 \circ a = a\cdot \Pi_e\q \forall a\in A \,.
\ee

The consequence of this construction is that if $A$ had
a Lie (super)algebra $l$ associated with some $t_i\in A$, then $M(A)$ admits
a symmetry algebra $l\oplus l$ generated by $t_i$ and $t_i\cdot \Pi_e$ with $t_i$
generating the diagonal subalgebra $l\subset l\oplus l$. This fact may have  important
implications for the space-time interpretation of the string-like HS theories
discussed in the next section.

\section{Interpretation}
\label{interpretation}

\subsection{Strings and tensor models}
Coxeter HS equations and their multi-particle extensions  have a number of features indicating their
relation to string-like models and their further tensor-like extensions anticipated to be
 holographic duals of the boundary  tensor sigma-models
 considered in \cite{Klebanov:2016xxf,Beccaria:2017aqc}.

First of all, the spectra of fields described by the rank-$p$ Coxeter HS models
with $p>1$ are far larger than of usual rank-one HS gauge theories. This is
most obvious from the fact that the zero-form fields $C(Y^n_\ga;k_v)$
depend on $p$ copies of the oscillators $Y^n_\ga$ as well as on the Klein
operators $k_v$ associated with all roots of the underlying Coxeter system
(modulo identification $k_{-v} = k_v$). This enlargement of the spectrum is in
qualitative agreement with the observation of \cite{Beccaria:2017aqc,Bulycheva:2017ilt} that
the spectrum of the boundary operators in tensor boundary models is far
richer than in the vector sigma-model. Most importantly, however,
the  Klein operators generating Coxeter reflections effectively permute the
arguments of the elementary master fields like $C(Y_1,\ldots Y_p;k|x)$.

In the absence of the Klein operators, products of
the fields would correspond to the tensor product of $p$ copies of star-product
algebras valued in $Mat_{N+M}$. However, in the process of solving Coxeter HS equations,
 relations of the type (\ref{kq})
will permute the variables $Y_n$ and $Y_m$ with different $n$ and $m$, not
affecting the matrix indices $a_n, b_n$ and $a_m, b_m$.
For instance, in the case of $p=2$, the star product of two master fields
$C(Y_1,Y_2|x) k_{12}$ gives
\be
\label{str}
(C(Y_1,Y_2|x) k_{12}) * (C(Y_1,Y_2|x) k_{12})  =
C(Y_1,Y_2|x) * C(Y_2,Y_1|x)\,.
\ee
As a result, nonlinear corrections to the $p=2$ system will contain products of
elementary strings  of master fields with repeatedly permuted arguments $Y_1$ and
$Y_2$
\be
C^n_{string}:=\underbrace{C(Y_1,Y_2|x) *C(Y_2,Y_1|x)*C(Y_1,Y_2|x)\ldots }_n\,.
\ee
 Such strings are analogous to the product of elementary matrix factors
and can be identified with the letters of an infinite Alphabet with
$n=0,1,\ldots \infty$. These are analogues of the single-trace operators
in the ordinary $AdS/CFT$ dictionary. General product of operators
decomposes into products of  elementary letters  analogous to
the multi-trace operators. For instance, operator (\ref{str}) as well as $C(Y_1,Y_2|x)$
are single-trace while $C(Y_1,Y_2|x)*C(Y_1,Y_2|x)$ is double-trace. Thus the spectrum  of
operators of the $p=2$ HS model is analogous to that of String Theory with the infinite set
of Regge trajectories. More precisely, to interpret these nonlinear combinations of
operators as corresponding to elementary states of some string-like model one has to
consider a multi-particle  HS model associated with the infinite set of elementary
$B_2$ systems
\be
  sym ( B_2\times B_2\times \ldots )\,,
\ee
where the graded symmetrization is  with respect to all elementary  $B_2$ factors in the sense
that all respective master fields $C(Y_1^1,Y_2^1,k_v^1;Y_1^2,Y_2^2,k_v^2; Y_1^3,Y_2^3,k_v^3;\ldots|x)$
are demanded to be graded symmetric  under the exchange of the variables $Y^a_1, Y^a_2, k^a_v$
associated with different factors of $B_2$ (\ie index $a$).

The $p=2$ Coxeter HS theory has deep parallels with the analysis of stringy
HS models by Gaberdiel and Gopakumar \cite{Gaberdiel:2014cha}-\cite{Gaberdiel:2017dbk}. In particular,
 the master fields of this theory depend on the two sets of oscillators $Y^A_{1,2}$ which is
close to saying that the stringy HS theory is based on two different HS symmetry algebras being
one of the conclusions of Gaberdiel and Gopakumar. It should be stressed that these algebras do not
commute with each other if the stringy coupling constant identified with the vacuum expectation value
of $F_{2*}(B)$ in (\ref{SScbb}) is non-zero inducing nontrivial Cherednik-like deformation (\ref{fqqc})
of the oscillator commutation relations. (Note that for conformal models considered by
Gaberdiel and Gopakumar the VEV of $F_{1*}(B)$ must be non-zero.)

Usual HS theory and its multi-particle extension result from the analogous construction applied
to $B_1$. Rank-$p$ tensor HS theories and their multi-particle extensions result analogously
from $B_p$ with $p>2$ as well as from other higher-rank Coxeter root systems. Clearly the pattern of
elementary operators of the $p>2$ higher-rank Coxeter HS models will increase enormously compared to
the $p=2$ string-like models which is  in agreement with  \cite{Beccaria:2017aqc,Bulycheva:2017ilt}.

To make the holographic correspondence more explicit it is necessary to compare
the spectrum of singlet boundary operators with that of the singlet sector
of the tensor bulk model. As explained in Section \ref{CHS}, due to
using framed  algebras,  these spectra do  match at least in the massless
sector  containing usual massless
states dual to the boundary currents in the holographic interpretation.
The  pattern of the full spectrum of the higher-rank Coxeter HS theories
 remains to be elaborated.

Let us stress that the Coxeter HS theories proposed in this paper are formulated in the
anti-de Sitter space
of appropriate dimension. In particular, this is true for the stringy $B_2$-HS models. Hence the
construction of this paper is different from that of genuine String Theory formulated in the
flat rather than, say, $AdS_{10}$ space. In fact, the reason why it is difficult
to formulate String Theory in $AdS_d$ is analogous to that discussed in Section \ref{hsmod}
for HS theory: a naive
attempt to deform the string spectrum to $AdS$ would immediately lead to infinite vacuum energy
since in this case all string modes have to contribute to the momentum generators to ensure that
their commutator is proportional to the Lorentz ones
\be
[P_n \,, P_m ] \sim -\Lambda M_{nm}
\ee
that act on all modes.
(Recall that the usual string
theory momentum operator is built from the zero modes \cite{Green:1987sp} which is not possible in
$AdS$.) Hence, the extension to the framed oscillator algebra can also be crucial to
reach a string-like theory in $AdS$.

An important related  feature of the $B_p$--HS models with $p\geq 2$,
is that they have two independent coupling constants
 instead of one in the usual $B_1$-HS theories.  These are the coefficients $\eta_{1,2}$
 in the linearized parts of the functions $F_{1,2*}(B)$ in (\ref{SScb}). The function $F_{1*}$
 is analogous to that of the rank-one (\ie $B_1$) HS theory. The respective term is important for the proper
interpretation of the equations in $AdS_4$ leading to the so-called central on-shell theorem
of \cite{Ann} representing the unfolded equations for massless fields in $AdS_4$. The function $F_{2*}$
first appears in the rank-two stringy model and, containing the Klein operators that
permute different species of $Y$-variables, is responsible for the
appearance of single-trace-like  strings of operators. The presence of two different
coupling constants is anticipated to have important implications for establishing
relation with usual string theory in flat space. One option is that to reach the latter
theory one has to take the limit with $\eta_2/\eta_1\to \infty$ which in turn may select
String Theory in critical dimension.

It should be stressed that it is a distinguishing property of the $B_p\sim C_p$
Coxeter group that  the related Cherednik system has two types of coupling constants
responsible for HS and stringy effects. It is not clear, in particular, whether
the $A_p$ and $D_p$ systems admit a meaningful HS interpretation since, due to the absence
of usual Klein operators reflecting signs of different species of oscillators $Y_A^i$,
the free field equations unlikely have a room for massless fields.

The full-fledged string theory is conjectured to be related to a rank-two multi-particle
$B_2$ Coxeter HS model. As explained in Section \ref{frames} following
\cite{Vasiliev:2012tv}, the form of the multi-particle
algebra significantly depends on the chosen basis. It remains to be analysed to which
extent the formulations in different frames are equivalent to each other. More precisely,
different frames of the Coxeter HS models of finite rank are equivalent.
 However, for the multi-particle  extensions the respective frame changes are in the
 infinite-dimensional space and  different frames may not be equivalent.  The details of
 description of multi-particle algebras in different frames are presented in Section \ref{frames} where the
 construction of \cite{Vasiliev:2012tv} is extended to the graded-symmetric case expressing the
 Pauli principle.

 This question
 may be related to the fundamental issue of the breaking of HS symmetries in the
 Coxeter HS theory.
Indeed, it is plausible to expect \cite{Caron-Huot:2016icg,Sever:2017ylk} that spontaneous breaking of HS
symmetries resulting in the appearance of massive HS fields is only possible in string-like models with the
infinite number of Regge trajectories. As conjectured in this paper the appropriate Coxeter HS theory is the
$B_2$ multi-particle theory. The goal is to break down HS symmetries to usual space-time symmetry of $AdS$ or
Minkowski type. The simplest way to do so is to let a rank-two zero-form topological field $B(Y_1;Y_2)$
 acquire a non-zero VEV
 \be
 B_0 =  Y_{i A} \cdot Y_j^A(\gga \sigma_1^{ij} +\rho \delta^{ij})\,,
 \ee
 where the Pauli matrix $\gs_1^{ij}$  and $\delta^{ij}$ are the two symmetric matrices
 invariant under the exchange $1\leftrightarrow 2$ in $i,j$.
 (Note that this proposal is somewhat reminiscent of the Girardello-Porrati-Zaffaroni
 mechanism \cite{Girardello:2002pp}.)
 Such $B_0$ preserves the $AdS$
 symmetry but breaks down the HS one. Note that so defined  $B_0$ is
 nonzero because elementary oscillators $Y^{iA}$ are anticommuting with respect to the  product
 $\cdot$ (\ref{grcom}).
  It remains to be seen how this VEV would affect (deform) the structure
 of the $AdS$ modules (field equations) of the originally massless
 fields. Spontaneous symmetry breaking would correspond to the mixture between the originally massless
 rank-one particle module and the rank-two current module. In the infinite-dimensional case of
 the multi-particle algebra the mechanism of such mixing may depend on the choice of the frame in
 the multi-particle algebra. As argued in Section \ref{framegen}, it is natural to anticipate
 that the proper frame is
 defined by (\ref{fus}), being associated with the boundary current algebra of \cite{Gelfond:2013xt}.
 Detailed analysis of this issue is one of the most urgent problems on the agenda.

\subsection{Space-time metamorphoses}
\label{met}
The formulation of HS equations in the unfolded form expressing the space-time exterior
derivative $\dr$ via the values of other fields like in equations (\ref{dW})-(\ref{dS})
on the fields $W$, $B$ and $S$ subjected to the constraints (\ref{SB}) and (\ref{SS})
allows us to unify in the same framework the systems that live in space-times of different
dimensions \cite{Gelfond:2010pm,Vasiliev:2012vf,Vasiliev:2014vwa}. This is achieved by letting the de Rham derivative $\dr$ be defined in the
infinite-dimensional space-time. Then the physical space-times like Minkowski or $AdS$ appear
when the background (vacuum) connection $W_0(x)$ of the respective symmetry group $G$ is
nontrivial with non-degenerate frame-like components along $x^\un_\|$ associated with
the translation (or
transvection) generators. In that case the  coordinates $x^\un_{\parallel}$
are observable  while the rest ones $x^\un_{\perp}$ are not:
 in the absence of components of forms   $W_0(x)$ along $x^\un_{\perp}$
the respective unfolded equations treated perturbatively would imply that, locally, all other
differential form fields in the system are either  $x^\un_{\perp}$-independent zero-forms
or pure gauge (\ie exact) $p>0$--forms. By this mechanism,  the system formulated
in the infinite-dimensional space-time is visualized by the coordinates $x^\un_{\parallel}$
representing one or another $G$--invariant space described by $W_0(x^\un_{\parallel})$.

In \cite{Gelfond:2010pm} it was observed that the higher-rank fields in lower dimension
can be interpreted as elementary fields in higher dimension. In the framework
of Coxeter HS theory this phenomenon acquires a direct realization as we
 explain now. Consider a spinorial Coxeter HS theory. In this case, the rank-one sector
 consists of the fields depending on a single  spinor variable $Y^A$ with $A=1,\ldots M$
 ($ M=2$ in the $3d$ model and $M=4$ in the $4d$ model). Bilinears of $Y^A$ form  generators
 of $sp(M)$ with respect to the star product. This is the $AdS_4$ algebra $sp(4)$ in the
  $4d$ model and a $sp(2)$ half of the $AdS_3$ algebra in the $3d$
model (to be doubled via introducing the Clifford element $\psi_1$ \cite{prok}). The diagonal
embedding of the  star-product generators of $sp^{diag}(M)$ into a rank-$p$ system
is
\be
\label{TAB}
t_{AB} = \sum_{i=1}^p \{Y_{Ai} \,,Y_{Bi}\}_*\,.
\ee
Vacuum connection $W_0(x)$ that describes usual space-time geometry like $AdS_4$ or $AdS_3$
is a flat $sp(M)$ connection ($sp(2)\oplus sp(2)$ in the $3d$ case).

Let $\Omega =A,i$. Operators
\be
T_{\Omega\Lambda} = \{Y_\Omega\,,Y_\Lambda\}_* *I_1*\ldots *I_p
\ee
are generators of $sp(p M)$. The diagonal embedding $sp^{diag}(M)$  into $sp(pM)$
is realized by the generators
\be
t'_{AB} = t_{AB}*I_1*\ldots *I_p\,.
\ee
The same time the $sp(M)$ generated by $t_{AB}$ (\ref{TAB}) acts diagonally on
the full framed Cherednik algebra.
Evidently,
\be
\label{id}
(t_{AB}-t'_{AB})*T_{\Omega\Lambda} =T_{\Omega\Lambda}* ( t_{AB}-t'_{AB})=0\,.
\ee

Let $\tau_{\Omega'\Lambda'}$ form a basis of $sp(pM)/sp^{diag}(M)$. Due to
 (\ref{id}), it is possible to identify the space-time components of the flat
connections of $t_{AB}$ and $t'_{AB}$ demanding them to have non-zero components
along coordinates $x^\un_\|$ of the space-time
$\M$ originally associated with the $sp(M)$ symmetry. Now it is possible
to choose a flat connection $W'_0(X)$ on some $sp'(pM)$-invariant space-time $\M'$ with local
coordinates $X$ in such
a way that the pushforward of $W'_0(X)$ to $\M$ gives $W_0(x)$.
This makes it possible to treat  the rank-$p$ fields in $\M$ as elementary fields in $\M'$.

One can proceed analogously with the additional species of oscillators in the
multi-particle algebra construction. From the analysis of Section \ref{uid} it
follows that the multi-particle $B_2$-HS model has $sp(8)\oplus sp(8)$ as
a finite-dimensional symmetry. The natural homogeneous space that admits this
symmetry is the group manifold $Sp(8)$ which is  $36$-dimensional. Moreover,
the conformal-like symmetry of $Sp(8)$ is $Sp(16)$ \cite{Vasiliev:2001zy,Didenko:2003aa,Plyushchay:2003gv}.

This has an interesting consequence that the usual space-time
interpretation of the Coxeter HS theories and their multi-particle extensions
is likely to have maximal  Minkowski dimension ten. This follows from the analysis
of the $sp(2M)$ invariant theories in \cite{Vasiliev:2001dc,Bandos:2005mb}
where it was shown how usual Minkowski space-time dimensions emerges
from the  $sp(2M)$-invariant equations with the conclusion that the
 maximal Minkowski space known to emerge from the $sp(16)$-invariant equations
 is ten dimensional. Though it is
tempting to conjecture that this is how the Superstring dimension ten emerges from the
Coxeter HS theory, details of this phenomenon need further investigation.
In particular, it would be interesting to see whether
this phenomenon is related to the twistor-like transform in ten dimensions introduced by Witten in
\cite{Witten:1985nt}.

Note that the mechanism explained in this section is based on the extension of
the space-time symmetries due to appearance of the additional species of oscillators in the
higher-rank Coxeter HS theories and/or  multi-particle extension of HS theories. As such it
is not quite the same as the holographic correspondence within the idempotent construction of
Section \ref{Proj} where the symmetry algebra $g$ remains the same but the $g$-module
pattern changes depending on the idempotent sector in question.

\section{Conclusion}
\label{concl}

We propose a class of Coxeter HS models conjectured to underly a symmetric phase
of String Theory (Coxeter group $B_2$) and its further tensor-like extensions
(Coxeter group $B_p$). These models contain two
coupling constants one of which is responsible for stringy effects (absent in the
conventional HS theory) and exhibit a number of interesting parallelisms with
String Theory. In particular, consistency of the holographic interpretation of
the boundary matrix-like model demands the latter to have $N=4$ SUSY.

The main idea of our construction was to find a formally consistent extension
of the known HS gauge systems \cite{more,prok,Vasiliev:2003ev} possessing a richer
spectrum and having a room for massless fields of all spins including spin-two
gravitational field. The former goal was reached via extension of the Coxeter
groups $Z_2$ or $Z_2\times Z_2$  generated by the Klein elements of the models of
\cite{more,prok,Vasiliev:2003ev} to any Coxeter group including the most important
case of $B_p$. To let massless fields be present in the model this construction was
further extended to the framed  algebras containing additional idempotent elements.

Though the main emphasize in this paper was on the spinorial HS models
somewhat analogous to Green-Schwarz superstring,
bosonic Coxeter HS theories of vectorial type considered in Section
\ref{vect} and their fermionic
counterparts to be elaborated  are also of interest as analogues of
the bosonic and fermionic strings. (For more detail see \cite{Vasiliev:2004cm}.)
We believe that  formal consistency in  presence
massless fields in the spectrum is so restrictive that it hardly leaves a
room for  consistent HS models beyond the list presented in this paper,
supplemented by the Coxeter extensions of to be constructed fermionic
generalizations of the model  of \cite{Vasiliev:2003ev}  in any dimension.
In particular it determines the structure of extended HS symmetries as well as
the field pattern.

Many of the important aspects of the proposed models, such as detailed analysis of
field spectra, spontaneous breaking of HS symmetries, holographic interpretation,
proper space-time interpretation, the choice of appropriate frame in the
multi-particle HS theories and  others  were only briefly sketched
in this paper, demanding a more detailed study delegated to the future as well as
some other issues including, for instance, the analysis of locality along the lines of \cite{GV,DGKV}.

An interesting open question is to give an interpretation to the Coxeter HS theories
based on the Coxeter groups different from $B_p$ which were of most interest in this
paper. In particular, it would be interesting to study more carefully the case of
Dihedral group $I_2(n)$ which is the symmetry group of the $n$-gone on the plane.
In this relation it should be noted that the case of $B_2$ is special due to
the isomorphism $B_2\sim I_2(4)$.
An interesting feature of the $B_2\sim I_2(4)$ HS theory is that, as shown recently in
\cite{Konstein:2017st,KT}, for certain
linear relations between the coupling constants $\nu_1$ and $\nu_2$
the respective Cherednik algebra acquires ideals. As a result,
on this locus of the plane of coupling constants some states in the system should
decouple. It would be interesting to investigate this phenomenon in detail,
especially in the context of unitarity.

\section*{Acknowledgments}
I am grateful to Nima Arkani-Hamed, Matthias Gaberdiel, Olga Gelfond, Igor Klebanov, Arkady Tseytlin, Herman Verlinde and
especially  Semyon Konstein for useful discussions.
This research was supported by the Russian Science Foundation Grant No 18-12-00507.

\end{document}